\documentclass[14pt, preprint2]{aastex62}
\graphicspath{{./}{plots/}}
\usepackage{float}
\makeatletter
\let\newfloat\newfloat@ltx
\makeatother
\usepackage{graphicx, subfigure, algorithm, algorithmic, amsmath}
\usepackage[normalem]{ulem}
\newcommand{\project}[1]{\textsl{#1}}

\newcommand{\apogee}{\project{\textsc{apogee}}}

\newcommand{\Gaiaeso}{\project{Gaia--\textsc{eso}}}
\newcommand{\galah}{\project{\textsc{galah}}}

\newcommand{\feh}{\mbox{$\rm [Fe/H]$}}

\newcommand{\xfe}{\mbox{$\rm [X/Fe]$}}
\newcommand{\alphafe}{\mbox{$\rm [\alpha/Fe]$}}

\newcommand{\rgal}{\mbox{$R_{\text{GAL}}$}}

\newcommand{\rbirth}{$R_\text{birth}$}
\newcommand{\tbirth}{t$_\text{birth}$}
\newcommand{\buckSim}{g7.55e11}

\newcommand{\buckHD}{g8.26e11}

\begin{document}
\title{Tracing birth properties of stars with abundance clustering.}

\author{Bridget L. Ratcliffe}
\affil{Department of Statistics, Columbia University, 1255 Amsterdam Avenue, New York, NY 10027, USA}
\author{Melissa K. Ness}
\affil{Department of Astronomy, Columbia University, 550 West 120th Street, New York, NY, 10027, USA}
\affil{Center for Computational Astrophysics, Flatiron Institute, 162 Fifth Avenue, New York, NY, 10010, USA}
\author{Tobias Buck}
\affil{Leibniz-Institut f\"{u}r Astrophysik Potsdam (AIP), An der Sternwarte 16, D-14482 Potsdam, Germany}
\author{Kathryn V. Johnston}
\affil{Department of Astronomy, Columbia University, 550 West 120th Street, New York, NY, 10027, USA}
\affil{Center for Computational Astrophysics, Flatiron Institute, 162 Fifth Avenue, New York, NY, 10010, USA}
\author{Bodhisattva Sen}
\affil{Department of Statistics, Columbia University, 1255 Amsterdam Avenue, New York, NY 10027, USA}
\author{Leandro Beraldo e Silva}
\affil{Jeremiah Horrocks Institute, University of Central Lancashire, Preston PR1 2HE, UK}
\author{Victor P. Debattista}
\affil{Institute of Space Sciences \& Astronomy, University of Malta, Msida MSD 2080, Malta}
\affil{Jeremiah Horrocks Institute, University of Central Lancashire, Preston PR1 2HE, UK}

\begin{abstract}

To understand the formation and evolution of the Milky Way disk, we must connect its current properties to its past. We explore hydrodynamical cosmological simulations to investigate how the chemical abundances of stars might be linked to their origins. Using hierarchical clustering of abundance measurements in two Milky Way-like simulations with distributed and steady star formation histories, we find that abundance clusters of stars comprise different groups in birth place (\rbirth) and time (age). Simulating observational abundance errors (0.05 dex), we find that to trace discrete groups of (\rbirth, age) requires a large vector of abundances.  Using 15-element abundances (Fe, O, Mg, S, Si, C, P, Mn, Ne, Al, N, V, Ba, Cr, Co), up to $\approx$ 10 clusters can be defined with $\approx 25$\% overlap in (\rbirth, age). We build a simple model to show that it is possible to infer a star’s age and \rbirth\ from abundances with precisions of $\pm$0.06 Gyr and  $\pm$1.17 kpc respectively. We find that abundance clustering is ineffective for a third simulation, where low-$\alpha$ stars form distributed in the disc and early high-$\alpha$ stars form more rapidly in clumps that sink towards the galactic center as their constituent stars evolve to enrich the interstellar medium. However, this formation path leads to large age-dispersions across the \alphafe-\feh\ plane, which is inconsistent with the Milky Way's observed properties. We conclude that abundance clustering is a promising approach toward charting the history of our Galaxy.
 
\end{abstract}

\section{Introduction} \label{sec:intro}

With large spectroscopic surveys, we have access to precise individual chemical element abundance measurements for 10$^5$-10$^6$ stars. The \galah\ survey \citep{Buder2018} provides stellar parameters and up to 23 abundances for 342,682 stars, and the \Gaiaeso\ survey \citep{Gilmore2012} measures detailed abundances for 12 elements in about 10,000 field stars. Another example of the current depth of observational data is the 16th data release of Apache Point Observatory Galactic Evolution Experiment (\apogee) survey, which contains information for 437,485 unique stars and more than 20 abundances \citep{apogeedr16, 2020AJ....160..120J}. 

For the Milky Way, we can use large spectroscopic surveys to catalogue an ensemble of measurements. These include precise stellar metallicities and abundances ([Fe/H], [X/Fe]) and imprecise ages, as well as current day positions and orbital parameters. These numbers can be used to work toward the reconstruction of the initial state of the Milky Way. While the chemical abundances of stars are birth properties, stars move over time and their dynamical properties change \citep[e.g.][]{2012MNRAS.426.2089R}. Chemical tagging utilizes the unchanging chemical abundances to identify star formation groups \citep{2002freeman-BH}. This is in theory possible as birth clusters up to $10^5 M_\odot$ are anticipated to be chemically homogeneous \citep{BH2010}. Chemical tagging has great promise \citep[e.g.][]{Hogg2016,2016martell}, however, it has been shown to be difficult due to the need for extremely large sample sizes \citep{Ting2015} and high precision data \citep{2013Lindegren}.  

In paper I of this series \citep{2020Ratcliffe}, we examined the distribution of clusters defined in a 19-dimensional chemical abundance space for 27,000 red clump \apogee\ DR14 stars in the Milky Way's disk \citep{bovy2014apogee}. Using a non-parametric agglomerative hierarchical clustering method, we determined that the groups defined in abundance space are spatially separated as a function of age.

Yet, to reconstruct the disk in the past, we need to know where the stars were born, which we can not measure directly from data. Recently, we have access to some of the largest and highest resolution samples of zoom-in Milky Way-like simulations (e.g. NIHAO-UHD, \cite{2020buck_NIHAO-UHD}; AURGIA, \cite{2017Auriga}; FIRE2, \cite{2018FIRE2}). We also now have access to more abundance information in simulations, with \buckHD\ from the NIHAO-UHD suite providing information for 15 abundances and Ananke from \cite{2020FIRE_sanderson} having information for 11 abundances. Thus, now in simulations, we have access to the full set of properties to trace formation and evolution of disks (e.g. [Fe/H], [X/Fe], age, and their origin as indicated by their birth radii within the galactic disk,  \rbirth). This enables us to use simulations to investigate and understand the dependencies and relationships between these properties in disk galaxies, under particular initial conditions and evolutionary events. 

Some work has examined \rbirth\ in simulations in an attempt to better understand the Milky Way's formation. With the use of an N-body simulation, \cite{2021Bird_avr} found that the age--\rbirth\ trends are a clear sign of the inside-out disc growth of their simulation, in addition to a correlation between \rbirth\ and birth kinematics. \cite{2021Johnson} find that in their hybrid hydrodynamical simulation there is a relationship between age, abundances, and \rbirth\ in the solar neighborhood, and that the low-$\alpha$ sequence represents a superposition of populations achieved by radial migration rather than an evolutionary sequence. See also \cite{2020_buckchemical}.

In this paper, we use simulations to explore the physical meaning of clusters of stars defined only in ([Fe/H], [X/Fe]) space in the observational data. The questions we wish to answer are (i) do abundances link to birth associations, and if so does it rely on star formation processes, (ii) how does the presence of observational errors and sample size effect results, (iii) are results dependent on clustering methods used, and (iv) is there a relationship between stellar birth properties and their abundances? Milky Way analogue simulations are a good tool to investigate the questions we pose and qualitatively represent formation processes of our Galaxy. Both observations and hydrodynamical simulations of Milky Way analogues show that from about z=1, stellar disks form inside-out, with on-going enrichment and star formation across the disk until late times \citep[e.g.][]{Haywood2013, 2020FIRE_sanderson}.

The numerical relationship between \rbirth, abundances, and age in the Milky Way has been investigated before. \cite{2018Minchev_rbirth} proposed a method to infer stellar birth radii from age and metallicity for observational data by projecting stars to the metallicity gradient corresponding to their age. Similarly, \cite{Frankel2018} invert the age--metallicity relation to find birth radii for the low-$\alpha$ sequence. \cite{Ness2019} also suggest that \feh, age, and high- or low-$\alpha$ sequence membership is all that is needed to infer a star's \rbirth. However, since \rbirth\ is unable to be measured observationally, none of these methods have been tested. Using simulations, we can study the link between chemical composition at birth ([Fe/H], [X/Fe]), birth time (\tbirth, or age) and birth location (\rbirth). 

This paper is organized as follows. In Section \ref{sec:data} we discuss the two simulations that this paper focuses on. Clustering methods used in this work are described in Section \ref{sec:methods}. Section \ref{sec:results_clustering} explores how chemically similar stars separate into discrete groups in the age--\rbirth\ plane, while Section \ref{sec:results_errors} explores how these results change under sampling and observational errors. Our last results section, Section \ref{sec:results_regression}, quantifies the relationship between (\feh, \xfe) and age, and (\feh, \xfe, age) and \rbirth\ using simple second order polynomial regressions. Finally, Sections \ref{sec:discussion} and \ref{sec:conclusion} present the discussion and conclusions of this analysis.

\section{Simulations} \label{sec:data}

\begin{figure*}
     \centering
     \includegraphics[width=.95\textwidth]{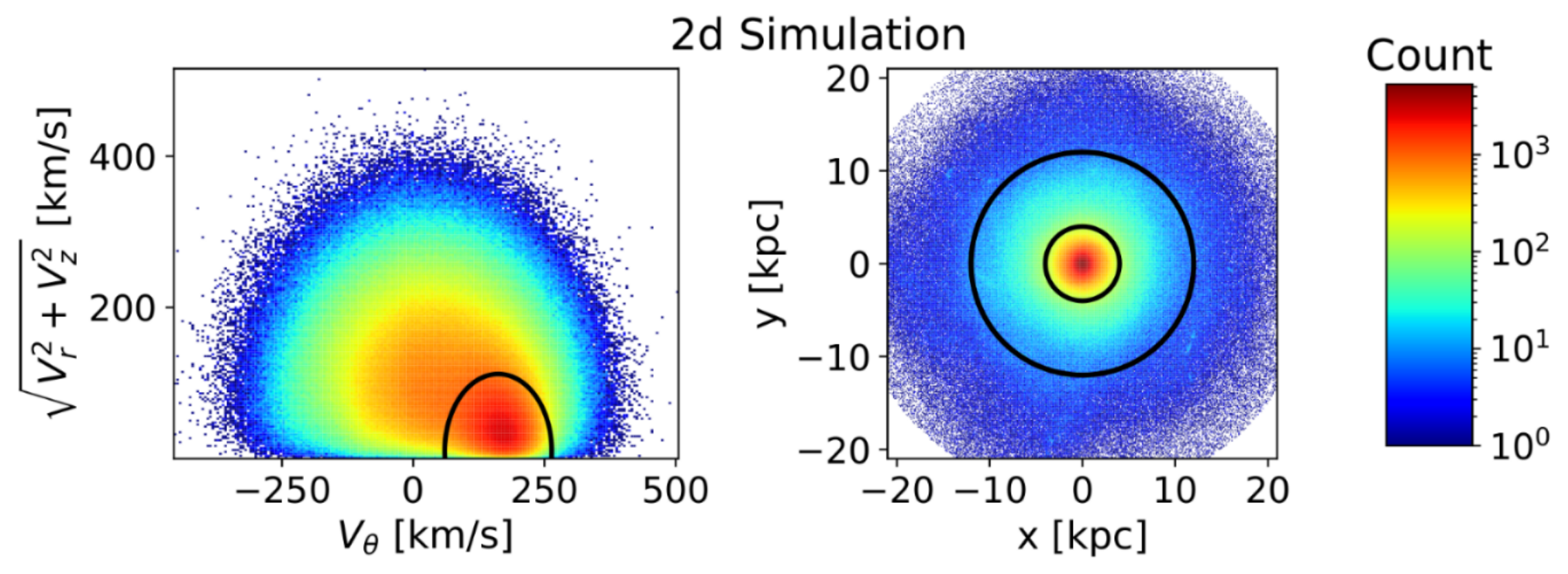}\\
     \includegraphics[width=.95\textwidth]{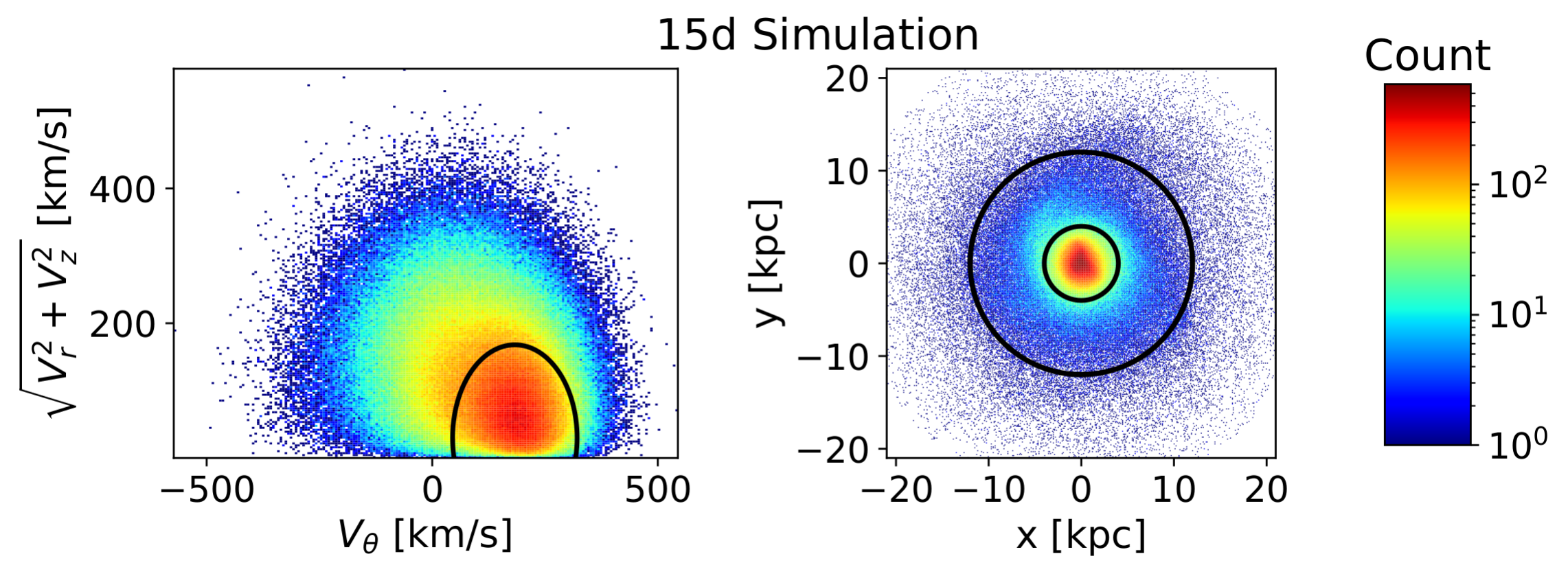}
\caption{Selection cuts of the (\textbf{top}) 2d, and (\textbf{bottom}) 15d simulations in the (\textbf{left}) velocity and (\textbf{right}) spatial planes. The black circle in the Toomre diagram marks where we defined the separation for the kinematically different disk and halo stars. Spatially, we define disk stars to have $|z|\leq 0.5$ kpc and $4 \leq$ \rgal\ $\leq 12$ kpc, shown in the right of the figure.}
\label{fig:selection_cuts}
\end{figure*}

In this paper, we focus on two simulations --- one with only two chemical abundances available (\buckSim, 2d) and another with 15 chemical abundances (\buckHD, 15d). We include both the high and low-dimensional simulations to investigate how stellar birth information is able to be captured in a few versus large vector of abundances.

The 2- and 15-dimensional simulations are taken from the Numerical Investigation of a Hundred Astronomical Objects (NIHAO) simulation suite \citep{2015MNRAS.454...83W}, and are part of the NIHAO-UHD suite \citep{2020buck_NIHAO-UHD,Buck2019}. The simulations were performed with the smooth particle hydrodynamics solver GASOLINE2 \citep{2017gasoline2}. They are both spiral disk galaxies with bulges, and were bulge dominated until redshift $z\geq 1$, with prominant stellar disks forming about 7-8 Gyr ago \citep{2020buck_NIHAO-UHD}. The 2-dimensional simulation has a stellar particle mass of 0.093 $\times 10^5 M_\odot$, while the 15-dimensional simulation has a stellar particle mass of 1.06 $\times 10^5 M_\odot$. For more details, see \cite{2020buck_NIHAO-UHD} for \buckSim\ and \cite{2021BuckHD_chemEnrich} for \buckHD. \footnote{The redshift zero snapshot and halo catalogue of the 2d simulation is publicly available for download here: \url{{https://tobibu.github.io/##sim_data}}. Additional files, e.g. the birth positions and higher redshift snapshots, as well as the 15d simulation snapshots can be shared upon request.}

There are two main differences between the simulations --- resolution and chemical enrichment prescription. As seen from the particle masses, the 2d simulation has a higher mass resolution while the 15d simulation is at fiducial resolution. However, NIHAO galaxies are numerically well converged, so for the purpose of this work the resolution difference should not matter. The other difference is a detail in the numerics. The 15d simulation has an updated chemical enrichment prescription as described in \cite{2021BuckHD_chemEnrich} which allows us to trace the 15 elements investigated here. This does not affect global properties of the galaxy such as stellar mass, star formation history, or disk size much.

One other difference is that the two galaxies have slightly different formation histories as they are two different realizations of $\Lambda$CDM initial conditions. This mainly affects the accretion history and the final stellar mass or disk size. However, what is important for this work is that the stellar disk properties of these simulations --- such as stellar mass, size, and rotation --- agree with observations of the Milky Way and local galaxies \citep{2020buck_NIHAO-UHD}. Furthermore, the age and \rbirth\ distribution in the \alphafe--\feh\ abundance plane is very similar to that observed in the Milky Way \citep[e.g.][]{yuxi2021universal, 2018Minchev_rbirth}.

The 2-dimensional abundance simulation has abundance information for \feh\ and [O/Fe]. The 15-dimensional simulation has the abundances of 15 elements from five different families --- five iron peak elements (Fe, V, Cr, Mn, Co), two light elements (C, N), two light odd-z elements (Al, P), five $\alpha$-elements (O, Mg, S, Si, Ne), and one s-process element (Ba). Both simulations show a bimodality in the \alphafe--\feh\ plane. The high- and low-$\alpha$ sequences in both simulations are a consequence of a gas-rich merger --- the high-$\alpha$ sequence evolves first in the early galaxy, while the low-$\alpha$ sequence forms after the gas-rich merger dilutes the interstellar medium's metallicity \citep{2020_buckchemical}. 

\subsection{Selection Cuts}\label{sec:selection_cuts}
We focus our analysis on the present day disk. We first select stars that overlap in space with the disk by imposing limits of  $|z| \leq 0.5$ kpc and 4 kpc $\leq$ \rgal\ $\leq$ 12 kpc, though are results are consistent for other spatial cuts. We then determine stars that are current disk members using 3-dimensional velocity space. We model a two component Gaussian mixture model in $(v_\theta, v_r, v_z)$ --- similar to the approach taken by \cite{2019ApJ...874...67B} and \cite{2018MNRAS.477.4915O} to model simulations in kinematic space using GSF \citep{2018ascl.soft06008O} --- and define disk stars to have a Mahalanobis distance less than 2 from the center of the corresponding Gaussian distribution. Figure \ref{fig:selection_cuts} shows these selection cuts in the equivalent Toomre diagram and $x-y$ plane for both simulations. Finally, we do an additional cut of $R_\text{birth} \leq 15$ kpc to ensure we are not looking at infalling debris.

Additionally, since the abundances cover different ranges, we make quality cuts on our abundance data in the scaled ([Fe/H], [X/Fe]) space, where the transformed abundances have mean 0 and a standard deviation of 1. Since the goal of this paper is to focus on global properties between abundances, age, and \rbirth, we remove outliers by only selecting stars that have scaled abundances between -4 and 4. Our final sample sizes are 229,045 and 44,359 particles for the 2- and 15-dimensional simulations respectively.

\subsection{Birth Properties in the Abundance plane} \label{sec:abundancePlane}
\begin{figure*}
     \centering
     \includegraphics[width=.97\textwidth]{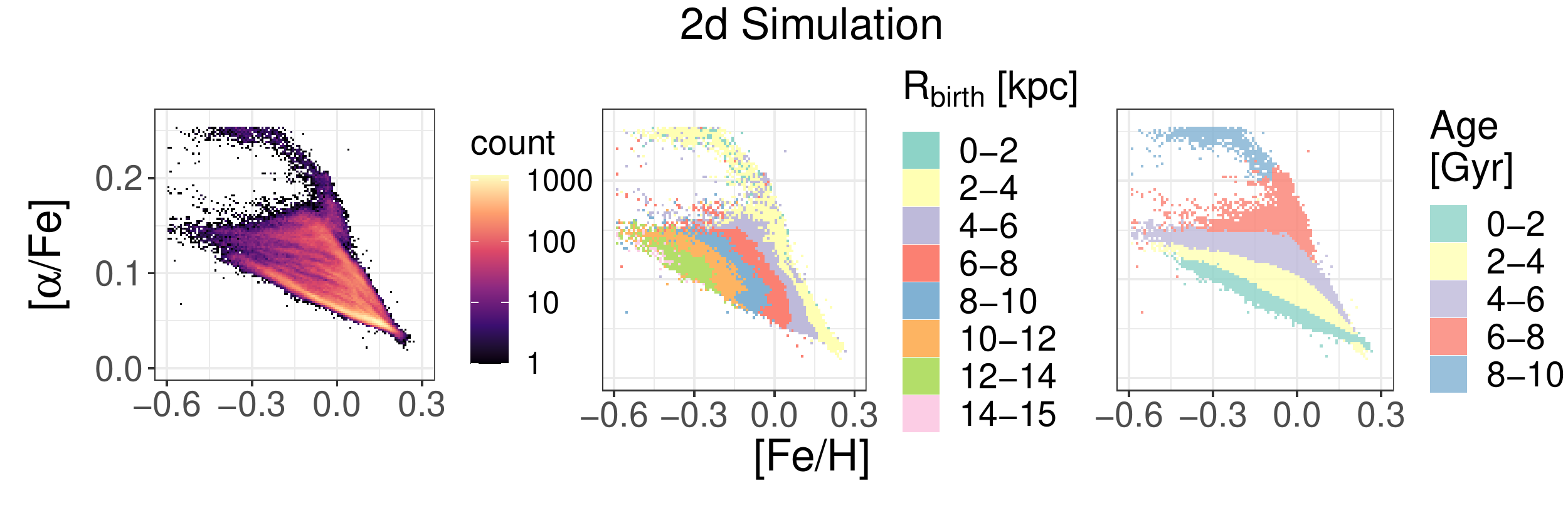}\\
     \includegraphics[width=1\textwidth]{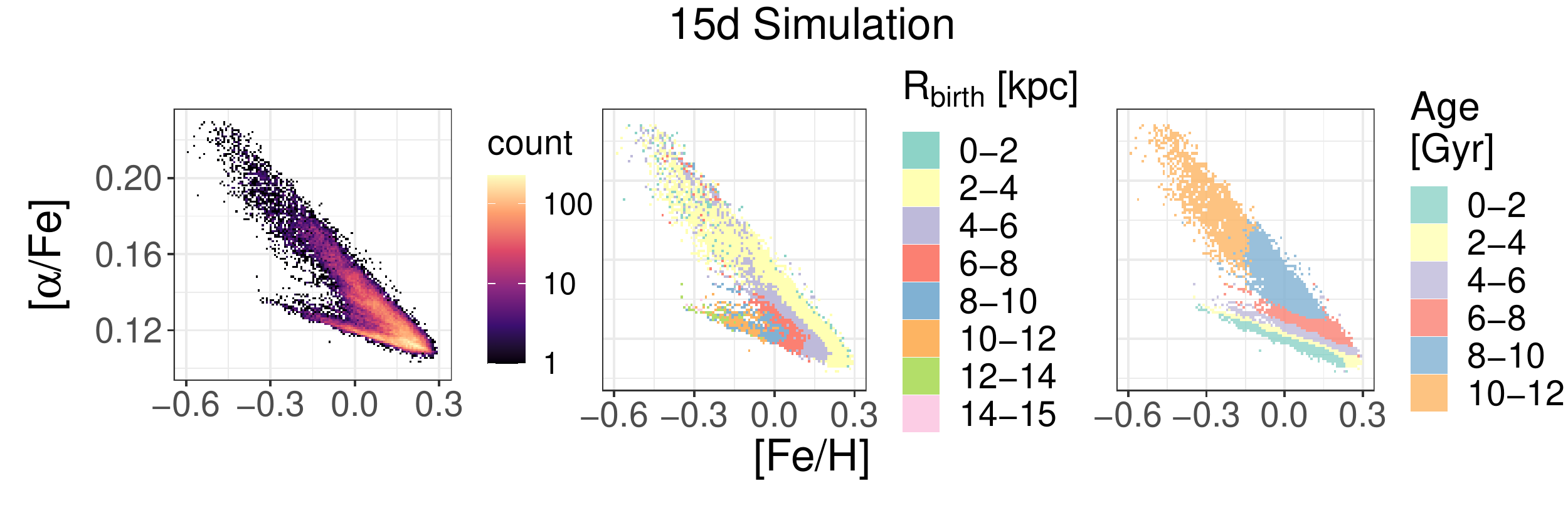}
\caption{The \alphafe--\feh\ plane of the (\textbf{top}) 2-dimensional, and (\textbf{bottom}) 15-dimensional simulation colored by (\textbf{left}) density, (\textbf{middle}) birth radius, and (\textbf{right}) age. The two simulations have a different footprint in this abundance plane (discussed in Section \ref{sec:data}), but both clearly have \rbirth\ and age trends due to their formation history.}
\label{fig:siml_3panel}
\end{figure*}

Figure \ref{fig:siml_3panel} shows the simulation data in the \alphafe--\feh\ plane after the selection cuts discussed above. Due to the formation history, both simulations have obvious trends in \rbirth\ and age (middle and right columns). For a given value of \alphafe, \rbirth\ increases as \feh\ decreases. Similarly, for a given value of \feh, age increases as \alphafe\ increases. As discussed in \cite{2020_buckchemical}, the horizontal age gradient and the diagonal radius separation in [$\alpha$/Fe]--[Fe/H] are simply a reflection of star formation in the disk happening at different radii, where metallicity decreases with increasing radii. 

The left column of Figure \ref{fig:siml_3panel} shows the density structure of the two simulations. For the 2-dimensional simulation (top left), there are density ridges which follow along different age tracks, whereas, there is no noticeable substructure in the 15-dimensional simulation (bottom left), presumably due to its lower mass resolution. The left panel of Figure \ref{fig:siml_3panel} also shows that the footprint of the 15-dimensional simulation is different than the 2-dimensional simulation. Most noticeably, the spread in this plane primarily captures high-$\alpha$ stars for the 15-dimensional simulation. The structural differences between the \alphafe--\feh\ planes of the 2d and 15d simulations are due to slight differences in their formation history (discussed above) and the different set of chemical yields for chemical enrichment (see \cite{2021BuckHD_chemEnrich} for discussion on the impact of yield tables and tracks in this abundance plane).

\section{Clustering Methods} \label{sec:methods}

We focus on two different clustering methods: agglomerative hierarchical clustering using Ward's minimum variance criterion \citep{ward1963hierarchical} and EnLink \citep{2009EnLink}. While both are nonparametric approaches, hierarchical clustering has the advantage of being simpler with only one tuning parameter --- the distance metric. On the other hand, EnLink needs two input parameters, but is able to fit complex structures since it has a locally adaptive distance metric. 

Unlike other clustering methods, hierarchical clustering and EnLink are nonparametric and thus do not force clusters to fit specific distributions. Additionally, other clustering methods, such as  K-means \citep{hartigan1979algorithm}, require prior knowledge for how many clusters comprise the data, whereas the two methods focused on in this work do not. Particularly in the high dimensional space of the 15d simulation --- where we cannot visualize all 15 abundance dimensions at once --- choosing the wrong number of clusters could give rise to misleading results for a method requiring the number of clusters beforehand.

\subsection{Hierarchical Clustering --- tree based clustering with a fixed distance metric}\label{sec:methods_hierarchical}
Following the same methodology \cite{2020Ratcliffe} used with observational red clump DR14 \apogee\ data, we use agglomerative hierarchical clustering using Ward's minimum variance criterion \citep{ward1963hierarchical} as one of the ways to combine the most chemically similar stars. Specifically, we use the Ward2 algorithm described in \cite{kaufman2009finding} and \cite{murtagh2011ward}. We conceptually describe the algorithm here, and refer the reader to \cite{2020Ratcliffe} for a more in depth explanation. 

The algorithm begins with each star as its own cluster, and at each step we combine the pair of clusters that leads to a minimum increase in total within-cluster variance until only one large cluster containing all the stars remains. The output is a tree showing the combination of groups at each step, called a dendrogram. Thus, the user decides the number of clusters to separate the data into after seeing the linking structure of the data.

\subsection{EnLink --- density-based clustering with an adaptive metric}\label{sec:methods_enlink}
EnLink is a nonparametric hierarchical clustering algorithm built on a locally adaptive distance metric and hence able to identify complex structures in the data. The full data set is first divided via a binary-partitioning algorithm which uses an entropy criterion to preferentially bisect dimensions that contain  maximum information \citep[``EnBid"][]{2006MNRAS.373.1293S}. This approach allows a nonparametric definition of ``local" regions in the data set from which the adaptive metric --- with flexible scales and orientations --- can then be derived (see \cite{2009EnLink} for full details). It is this metric that defines the distance between particles subsequently.

EnLink partners the adaptive distance metric machinery with IsoDen \citep{pfitzner1997halo}, which is a density-based clustering algorithm. Conceptually, clusters can be thought of as regions around high density peaks that are separated from one another by lower density regions. Thus, as we lower the isodensity contours when examining a high density region, we stay within the cluster until we encounter a lower density region that connects to another high density region. Then, as the isodensity contour continues to lower, a new group encompassing both clusters is formed. Continuing in this fashion forms a hierarchy of density-based parent-child clusters. 

EnLink has two user specified parameters: the number of nearest neighbors used in calculating density ($k_\text{den}$) and the threshold significance level when comparing the high and low density levels of parent-child clusters ($S_\text{th}$). Since the goal of our analysis is not to find the best clustering, but rather to investigate the stability of our results, we choose to vary $k_\text{den}$ from 30 to 1,000, and $S_\text{th}$ is such that the expected number of groups due to Poisson noise is 0.5, 1, and 2. We did not observe major differences in our results.

\section{Results I: Abundance clusters form groups of (\rbirth, age)} \label{sec:results_clustering}

\begin{figure*}
     \centering
     \includegraphics[width=1\textwidth]{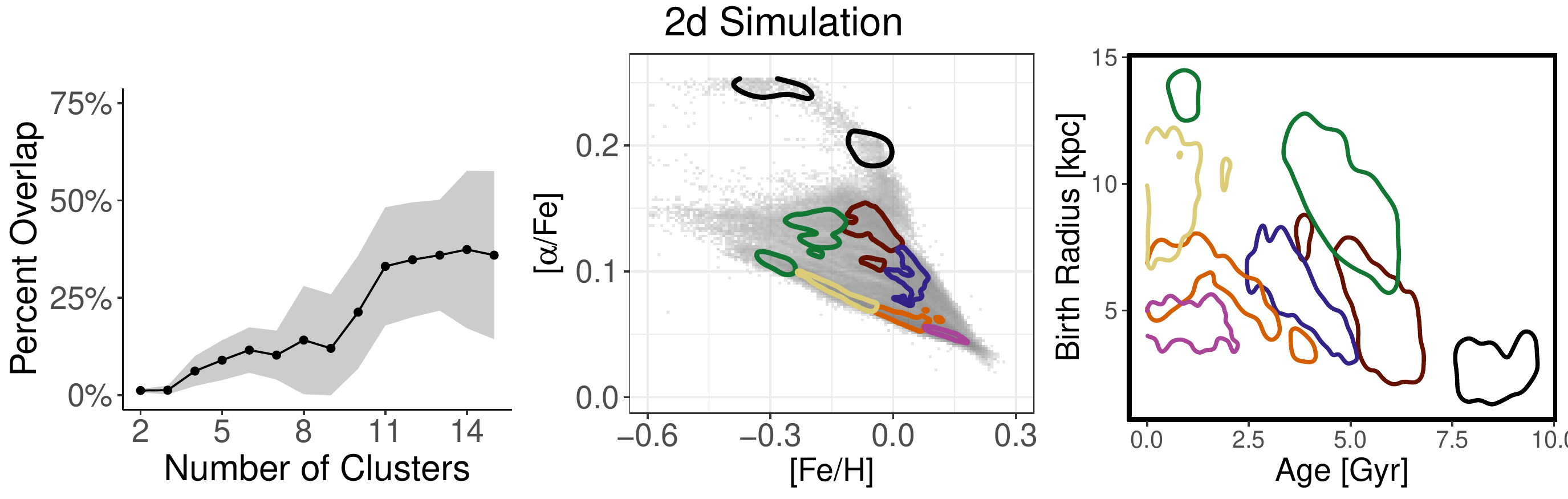}\\
     \includegraphics[width=1\textwidth]{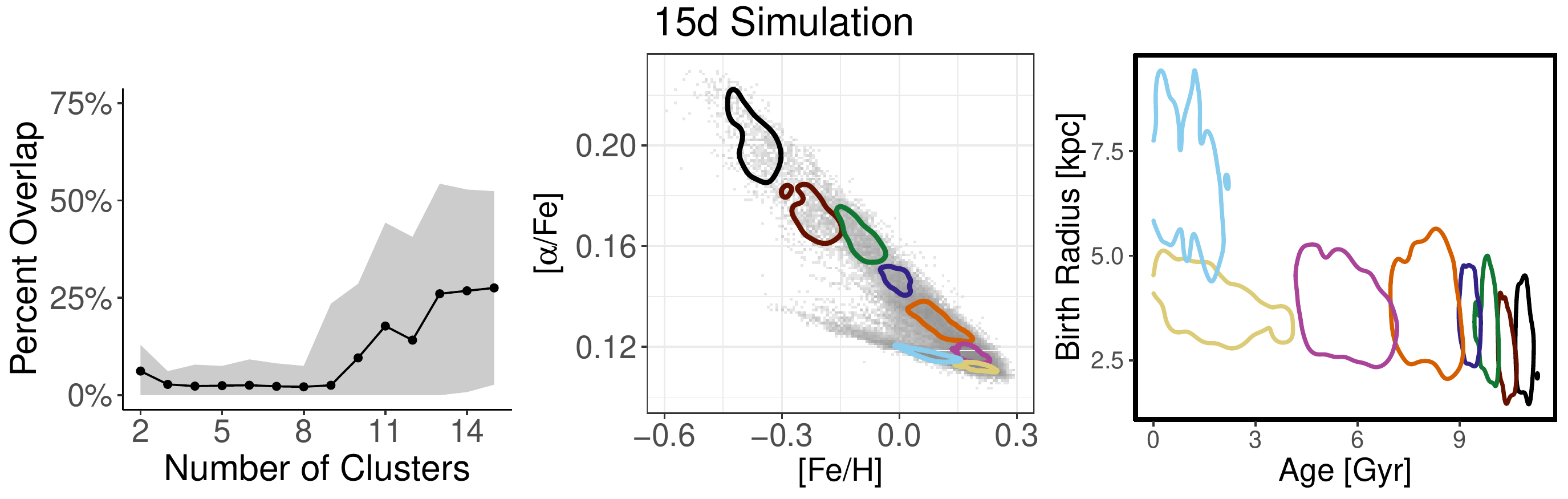}\\
     \includegraphics[width=1\textwidth]{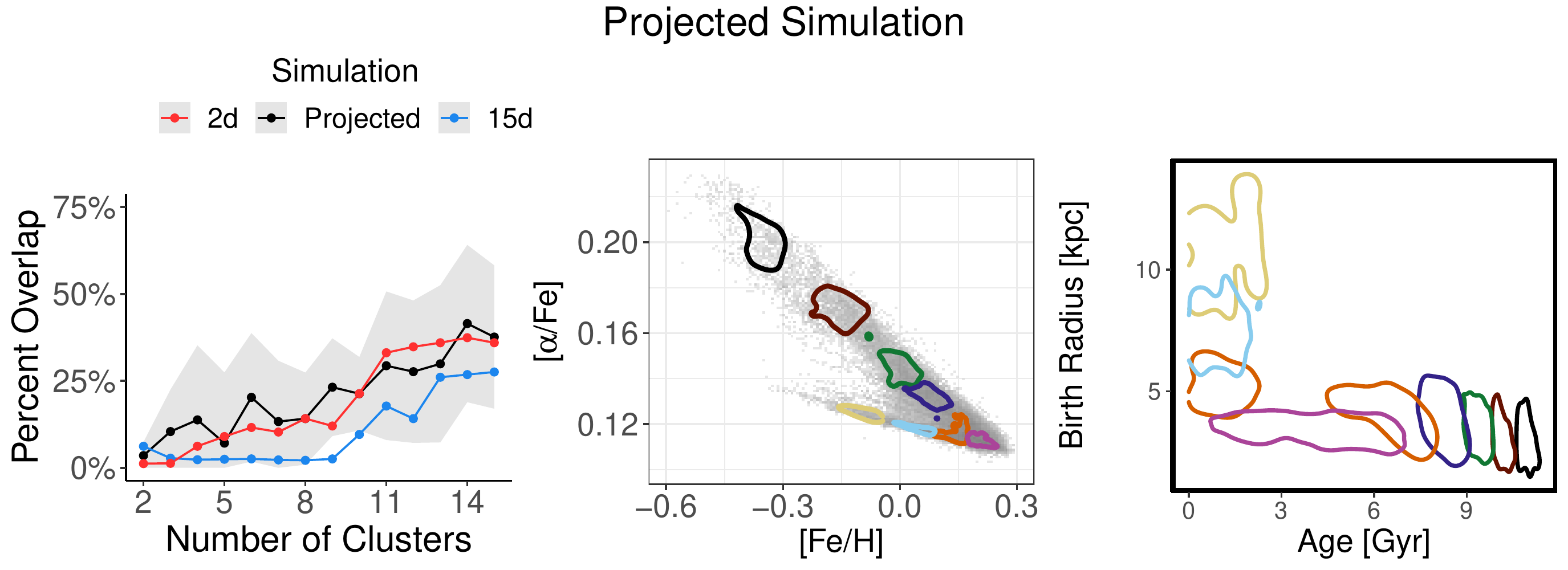}
\caption{(\textbf{Left}) The median overlap percentage for clusters in the age--\rbirth\ plane as a function of the number of clusters determined by hierarchical clustering for the (\textbf{top}) 2-dimensional, and (\textbf{middle}) 15-dimensional simulation.  Each point is determined by calculating the median percent each cluster overlaps with the other clusters at the 50\% contour level in the age--\rbirth\ plane. The grey ribbon represents one standard deviation about the median overlap percentages. The 50\% contour lines of clusters projected into the (\textbf{middle}) \alphafe\--\feh\ plane, and (\textbf{right}) age--\rbirth\ plane. There are seven and eight clusters for the 2- and 15-dimensional simulations respectively. The (\textbf{bottom}) row shows the results for clustering in only the \alphafe--\feh\ plane of the 15d simulation (labeled ``Projected") in comparison to clustering in the full 15-dimensional chemical space and the 2d simulation.}
\label{fig:contour_3panel}
\end{figure*}

In this section, we investigate the birth properties of clusters defined in a 2-dimensional (Section \ref{sec:results_2d}) and 15-dimensional (Section \ref{sec:results_tagsHD}) chemical abundance space. We examine the amount each cluster overlaps in birth time and space from 2 to 15 clusters determined using both the 2d and 15d simulations.

\subsection{Two Abundances Tag Distinct Ages and \rbirth}\label{sec:results_2d}

As mentioned in Section \ref{sec:methods_hierarchical}, hierarchical clustering produces a dendrogram showing how stars in abundance space combine, starting from each star being its own group to one cluster containing every star. After the linking structure is determined, the user then specifies how many clusters to separate the sample into. Walking down the tree --- and thus increasing the number of clusters --- corresponds to one group being separated to form two new clusters at each step. For a given number of $k$ clusters defined in the 2- or 15-dimensional abundance space, we determine the contour level that contains 50\% of the stars within each abundance cluster after projecting into the age--\rbirth\ plane. We then calculate the percent that each 50\% contour level overlaps with the other $k-1$ cluster's 50\% contour levels by laying down a fine grid and comparing the number of points in just the $i$th cluster to the number of points that fall in more than just the $i$th cluster. For the $i$th cluster, the overlap percentage is defined as the percent of area that is common between the 50\% contour region of the $i$th cluster and the 50\% contour regions of the other $k-1$ clusters. 

The top left panel of Figure \ref{fig:contour_3panel} shows the median of these overlap percentages and standard deviation of the $k$ overlap percentages as a function of the number of clusters found in the 2-dimensional simulation using hierarchical clustering. We see that clusters defined solely using two abundances show consistently low overlap in birth time and space for up to 10 clusters at the 50\% contour level (and up to seven clusters at the 75\% contour level, which is not shown). 

The middle and right panels of the top row in Figure \ref{fig:contour_3panel} show the seven clusters in the \alphafe--\feh\ and \rbirth--age planes at the 50\% contour level. We can see that the clusters found using two abundances separate diagonally, both as a function of age and \rbirth. We believe that this primarily is a consequence of formation history, as the low-$\alpha$ stars have gradients in abundances, age, and \rbirth.

\subsubsection{Using EnLink to Leverage Density Structure in the 2d Simulation} \label{sec:enlink_2d}
\begin{figure*}
     \centering
     \includegraphics[width=.85\textwidth]{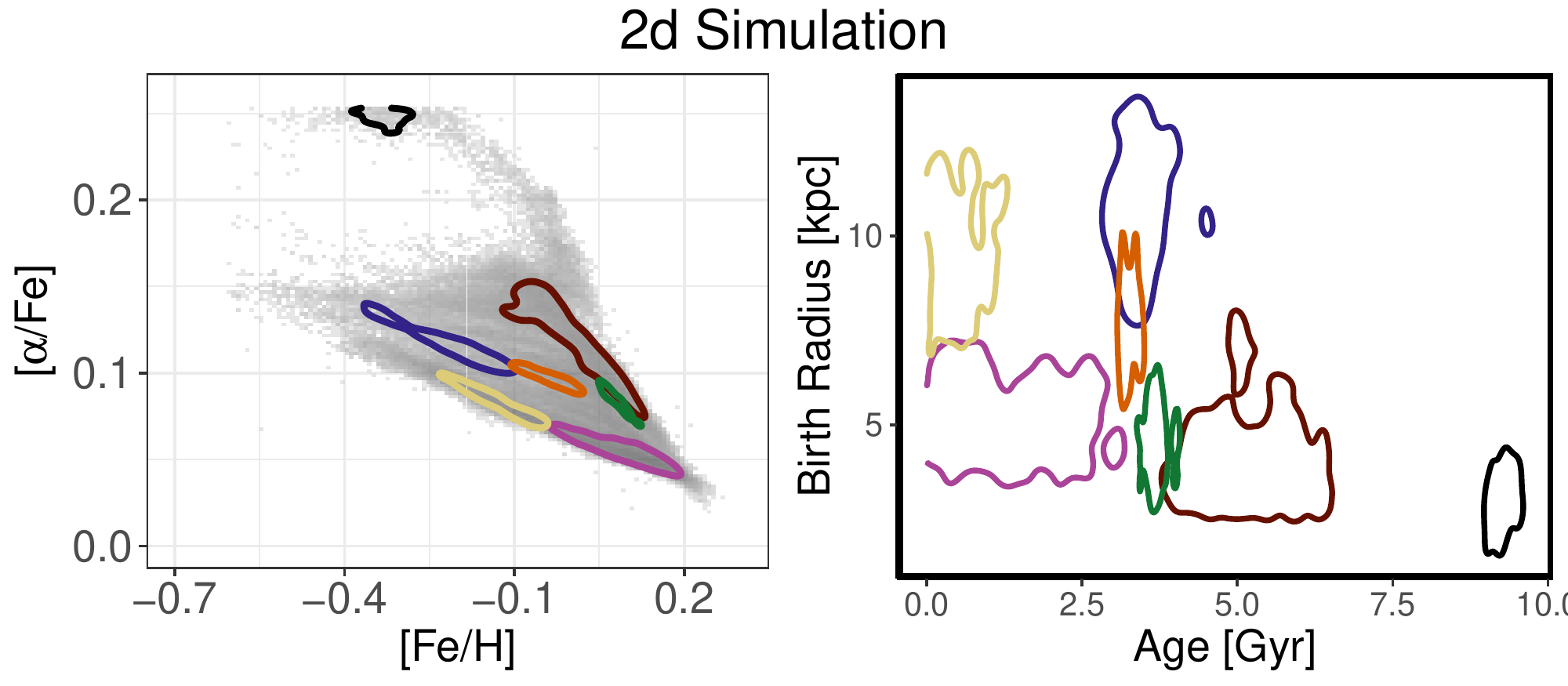}
\caption{Seven clusters found using the density based nonparametric method EnLink in the 2-dimensional abundance space of \buckSim\ projected into the \alphafe--\feh\ and age--\rbirth\ planes. All stars are assigned to a cluster, and each contour captures 50\% of the stars within the cluster. The clusters now follow the density ridges discussed in Section \ref{sec:abundancePlane} and show more separation in the age--\rbirth\ plane than those clusters found using hierarchical clustering. The median amount each cluster overlaps is 1\%.}
\label{fig:enlink_clusters}
\end{figure*}

In the 2-dimensional abundance space of \buckSim, we can see streaks of higher density regions along the different age bins (see top left of Figure \ref{fig:siml_3panel}). Hierarchical clustering does not leverage the density of the simulation in abundance space and thus is unable to capture the streak formations (see top middle of Figure \ref{fig:contour_3panel}). Therefore, to include this structure as information in assignment of cluster groups, we employ a density based clustering method with an adaptive distance metric to use the ridge-like structure in association of groups.

As discussed in Section \ref{sec:methods_enlink}, EnLink has the number of nearest neighbors used to determine the density at a point and the maximum number of spurious clusters created by noise as input parameters. We find that the cluster separation in the age--\rbirth\ plane is fairly stable when we focus on allowing either 0.5 or 1 spurious clusters and 30 to 1,000 nearest neighbors, with the majority of EnLink clusters having a median overlap percent of below 25\% at the 75\% contour level, and near 0\% at the 50\% contour level.

Figure \ref{fig:enlink_clusters} shows the 50\% contours in the abundance and age--\rbirth\ planes for seven clusters and parameter settings 260 nearest neighbors and a maximum of one spurious cluster. We can clearly see that EnLink captures the streaks of overdensity in the \alphafe--\feh\ plane better than hierarchical clustering, and in doing so we see better separation in the age--\rbirth\ plane, with a median overlap percentage of only 1\%. Since EnLink traces the higher density ridges, and the ridges trace different age bins, the cluster separation in the age--\rbirth\ plane no longer follows the diagonal trends given by hierarchical clustering. Each group primarily lives in a unique birth time and place, though there is some overlap with the middle aged clusters, possibly due to not fine tuning the algorithm. Overall we conclude that for the 2-dimensional high resolution simulation, using a density based method with an adaptive metric is desirable for the best results.

\subsection{Additional nucleosynthetic families and abundances provide more information about birth properties}\label{sec:results_tagsHD}

So far we have demonstrated that just two abundances (\alphafe\ and \feh) can trace separate ages and birth radii. We now investigate how additional abundances, including additional nucleosynthetic families, strengthens this result. The list of abundances and their families is given in Section \ref{sec:data}. Due to the difficult problem of estimating density in a high dimensional space, in addition to the issue of tuning the algorithm, we do not use EnLink to cluster in 15 dimensions, and instead focus on hierarchical clustering.

As shown in the middle left panel of Figure \ref{fig:contour_3panel}, clusters defined in the chemical space of 15 abundances show more separation in the age--\rbirth\ plane than the clusters defined in the abundance space of the 2-dimensional simulation. These exhibit separation for 13 clusters at the 50\% contour level, and eight clusters at the 75\% contour level (not shown). As shown in the middle and right columns of the middle row of Figure \ref{fig:contour_3panel}, the clusters comprised of older stars (which trace the high-$\alpha$ stars) primarily show separation as a function of age, whereas the younger low-$\alpha$ stars show separation in both \rbirth\ and age. This shows that high-$\alpha$ stars were all born near the Galactic center, whereas the low-$\alpha$ stars were born at different radii and times throughout the galaxy.

Comparing these results to the 2-dimensional simulation (top left of Figure \ref{fig:contour_3panel}), we see that clusters defined in the 15-dimensional abundance space using hierarchical clustering overlap less in age and \rbirth. In particular, for up to nine clusters, the 15-dimensional clusters are predominantly distinct in birth time and space at the 50\% contour level, whereas in 2-dimensions the clusters have some overlap even for as few as four clusters. 

\subsection{Comparing groups in 2d and 15d}

In the previous section (Section \ref{sec:results_tagsHD}) we showed that clusters defined in the 15-dimensional abundance space of \buckHD\ showed more separation in age and \rbirth\ than clusters defined in the 2-dimensional abundance space of \buckSim. Here we show the results of clustering in just the \alphafe--\feh\ plane of the 15-dimensional simulation are consistent to those of the 2-dimensional simulation. 

The bottom left of Figure \ref{fig:contour_3panel} shows that the amount abundance clusters overlap in the age--\rbirth\ plane is similar between the 2-dimensional simulation (red) and the \alphafe--\feh\ plane of the 15-dimensional simulation (black; labeled ``projected"). As the simulations are split into more clusters, both consistently show less distinction in (age, \rbirth). On the other hand, the clusters defined in the full 15-dimensional simulation retain separate ages and birth radii for up to nine clusters. This shows that more separate birth information is retained in abundance clusters when more abundances are included, and therefore we claim abundance clusters being more separate in age and \rbirth\ is due to additional abundance information, and not an artifact of different simulation history or resolution.

\begin{figure}
     \includegraphics[width=0.45\textwidth]{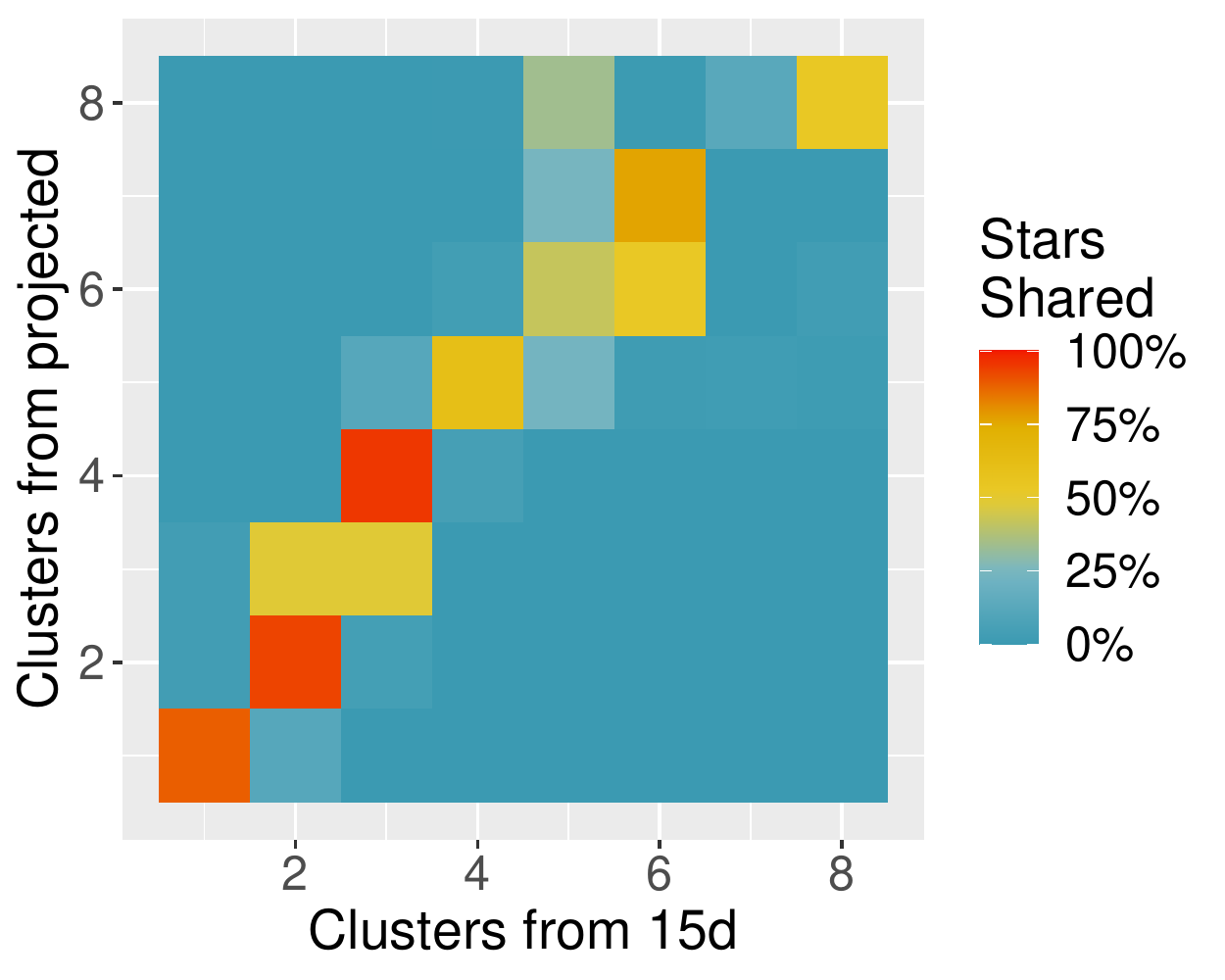}
\caption{Comparison of the number of stars shared between eight clusters defined in all 15 abundances versus just (\alphafe, \feh) of the 15-dimensional simulation. The clusters are arranged in order of age, with cluster 1 being the oldest and cluster 8 being the youngest. The percent of stars shared between clusters is determined by counting the number of stars found in the projected group that are in the 15-dimensional group, and then dividing by the number of stars in the projected group.}
\label{fig:15in2d_heatmap}
\end{figure}

Figure \ref{fig:15in2d_heatmap} compares eight clusters defined in the \alphafe--\feh\ plane of the 15-dimensional simulation to eight clusters defined in the full 15-dimensional abundance space. The clusters are arranged in order of age, with cluster 1 being the oldest and cluster 8 being the youngest. The oldest clusters share the most stars between the two simulations, whereas the middle aged and youngest clusters are more muddled.

\subsection{A grid in \alphafe--\feh\ separates birth properties less effectively}
\label{sec:results_whyCluster}

So far we focused on how chemically similar stars differ in birth time and space using clustering methods. Now we motivate why using a clustering method is needed for grouping stars.

\begin{figure*}
     \centering
     \includegraphics[width=0.47\textwidth]{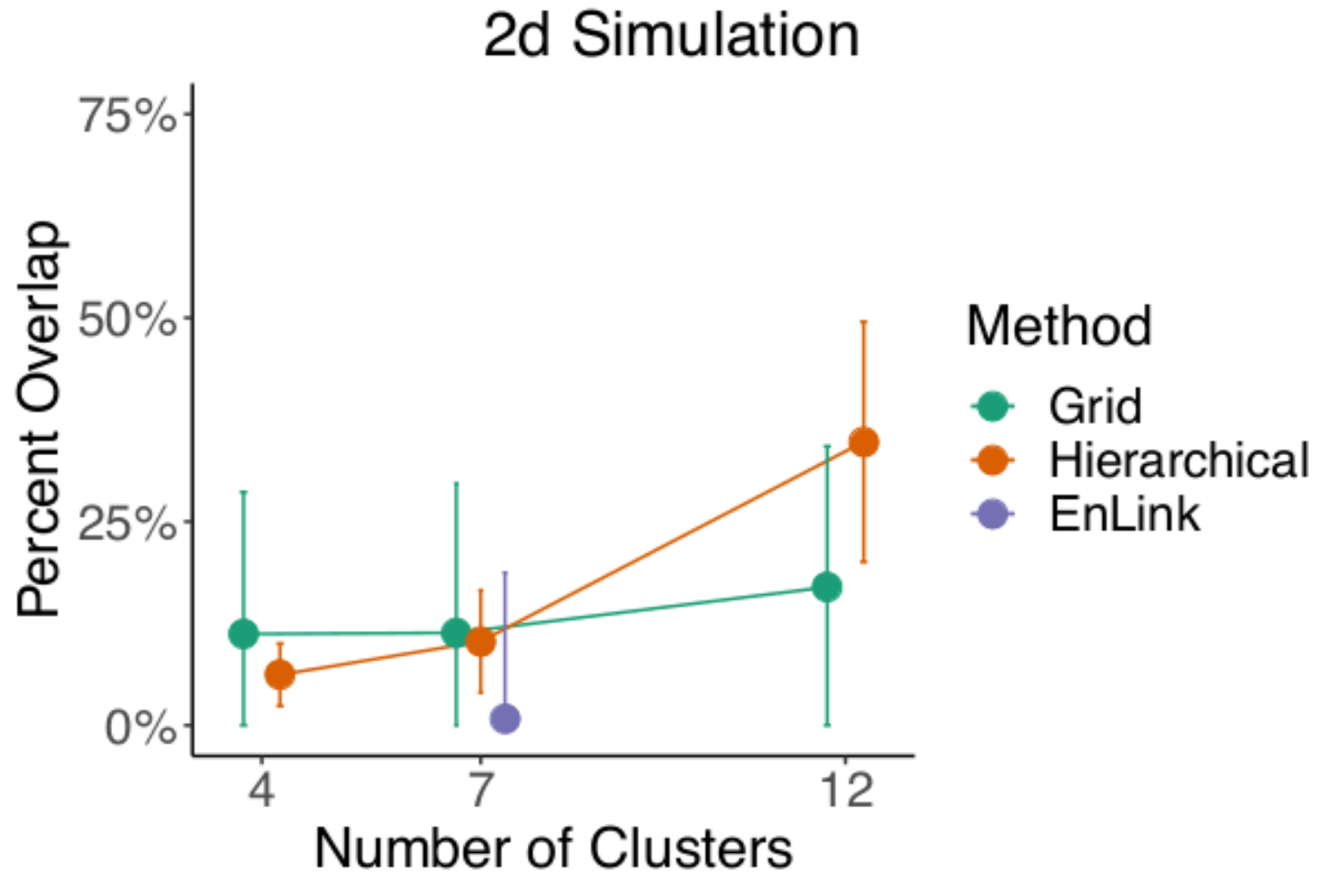}
     \includegraphics[width=0.47\textwidth]{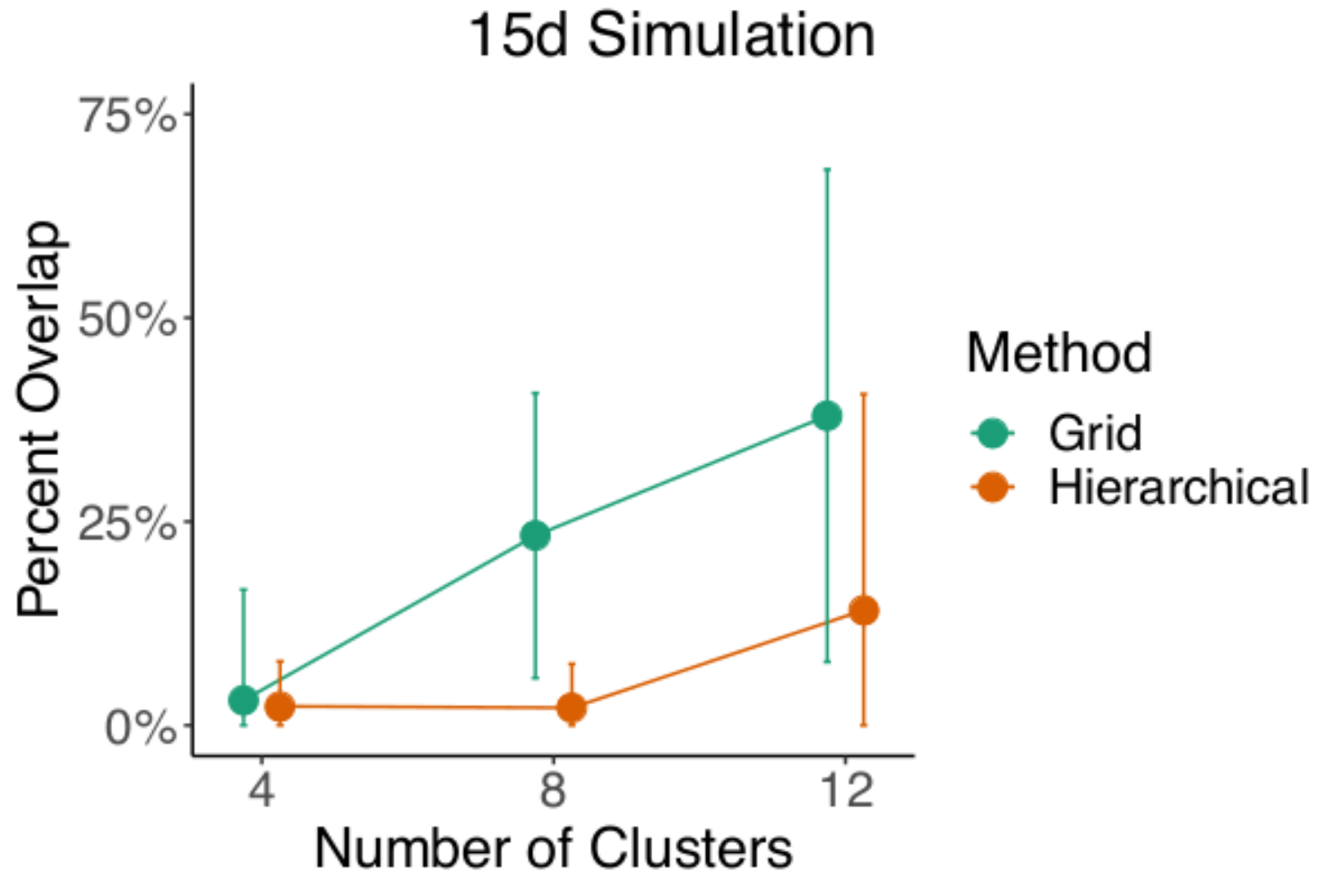}
\caption{Each point and error bar represents the median and standard deviation of the amount clusters overlap with one another in the age--\rbirth\ plane. Comparing three different ways of combining stars in chemical space (a grid laid out in the \alphafe--\feh\ plane, hierarchical clustering, and EnLink) shows that if we only have 2 abundances available (\textbf{left}), then groups of chemically similar stars determined via hierarchical clustering and gridding on average show similar separation in the age--\rbirth\ plane. Leveraging the density in the abundance plane allows for even better separation in birth time and space. When we include more abundances and nucleosynthetic channels (\textbf{right}), we find that hierarchical clustering done in 15 dimensions yields more distinct groups in age--\rbirth\ than gridding in the visual two dimensions. Note that EnLink was inconclusive in 15 dimensions due to the curse of dimensionality and also only reported seven clusters for the 2d simulation, and therefore no EnLink result is shown for the 15-dimensional simulation or for four and twelve clusters for the 2d simulation.}
\label{fig:clusterVgrid}
\end{figure*}

Figure \ref{fig:clusterVgrid} shows the median and standard deviation of the percent cluster overlap for both the 2-dimensional and 15-dimensional simulations when separating the stars using different grouping methods. We compare hierarchical clustering with laying down a Cartesian grid in the \alphafe--\feh\ plane. The number of \alphafe--\feh\ bins are chosen to produce four, seven/eight (for 2/15d simulations respectively), and twelve clusters. We also show the results of EnLink for the 2-dimensional simulation. The left panel of Figure \ref{fig:clusterVgrid} shows that when only two dimensions of abundance information are known, we do not gain any more knowledge of birth properties from using a simple clustering method than if we were to create bins by laying down a Cartesian grid across the \alphafe--\feh\ plane. The clusters retain more separate birth properties when leveraging density with an adaptive distance metric, however the errors are higher than that of binning and hierarchical clustering. 

The right panel of Figure \ref{fig:clusterVgrid} reveals that when higher dimensional abundance information is available, there is a noticeable difference between only looking at stars separated using a grid in \alphafe--\feh\ versus using a clustering method in the full abundance space. Not only does clustering in 15 dimensions produce less overlap in the age--\rbirth\ plane, but based on the smaller standard deviation, groups defined using hierarchical clustering produce consistently small overlap between clusters whereas groups defined by binning in the \alphafe--\feh\ plane produce a large range of amount clusters overlap in birth time and place.

\begin{figure*}
     \centering
     \includegraphics[width=.95\textwidth]{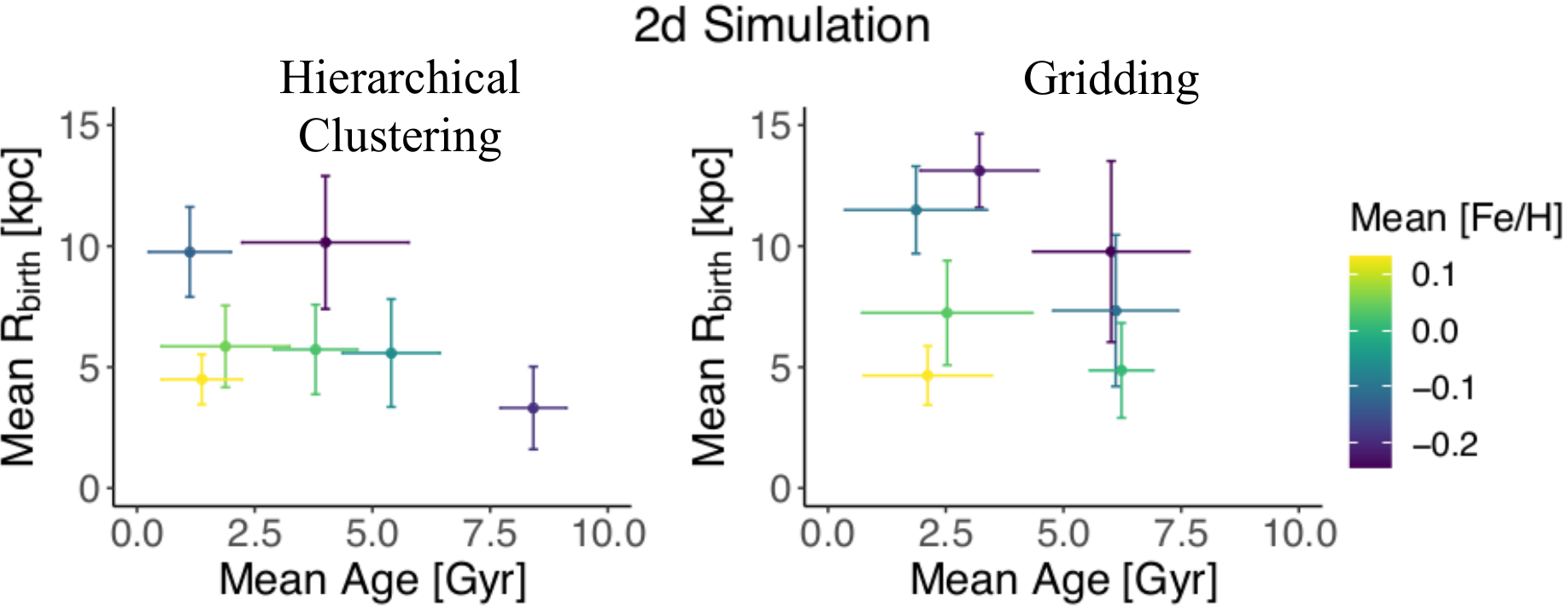}\\
     \includegraphics[width=.95\textwidth]{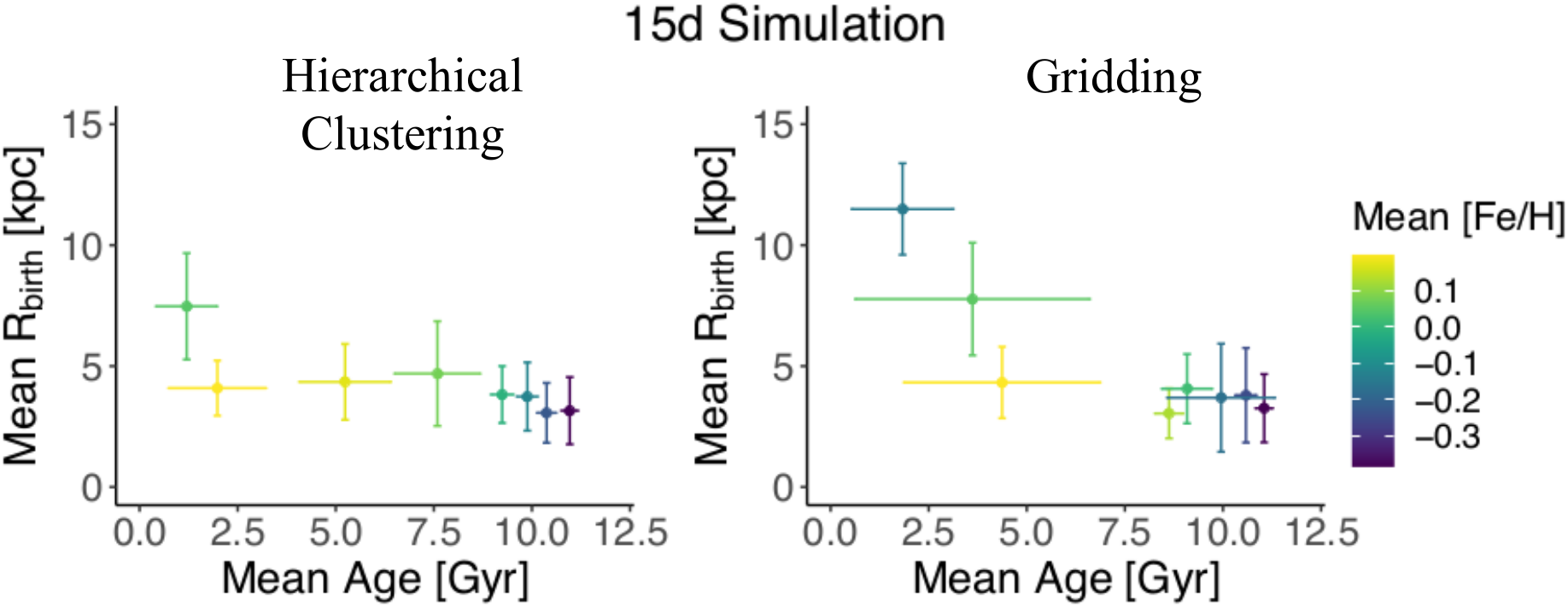}
    \caption{Seven and eight clusters separated using (\textbf{left}) hierarchical clustering (\textbf{right}) and gridding in the \alphafe--\feh\ plane for the (\textbf{top}) 2-dimensional and (\textbf{bottom}) 15-dimensional simulation respectively projected into the \rbirth--age plane. Each point represents the mean (age, \rbirth) for each cluster, colored by the mean metallicity. Error bars shown are 1-$\sigma$ standard deviations. Groups defined using hierarchical clustering show a metalicity gradient for a given \rbirth, suggesting that the clusters are physically meaningful. Groups defined in mono-\alphafe--\feh\ bins do not show a metalicity gradient, and have larger dispersions in age.}
\label{fig:meanRbirthVage_feh}
\end{figure*}

Figure \ref{fig:meanRbirthVage_feh} compares the mean and standard deviation of (age, \rbirth) of each cluster determined using two different methods: hierarchical clustering (left) and binning in the \alphafe--\feh\ plane (right) to create a cartesian grid. This figure demonstrates that clusters defined using hierarchical clustering preserve physical interpretation and agree with expected trends, namely we see a clear metallicity gradient as a function of age for a given \rbirth\ when stars are separated using hierarchical clustering. The trend is not as obvious to see when stars are separated in the \alphafe--\feh\ plane with a grid. Additionally, we see that clusters defined using a clustering method tend to have less overlap, and the overlap they do have is consistent among nearly all the clusters. However for the groups defined via a grid, the overlap between clusters is irregular, with some groups being mainly separate and others completely overlapping multiple groups.

Additionally, particularly for the 15-dimensional simulation, the age dispersion for each cluster when defined using hierarchical clustering is much smaller than when clusters are separated using a grid. This again shows that the groups found using hierarchical clustering represent different physical groups in time.

\section{Results II: Clustering in Observational Data} \label{sec:results_errors}

\begin{figure*}
     \centering
     \includegraphics[width=.88\textwidth]{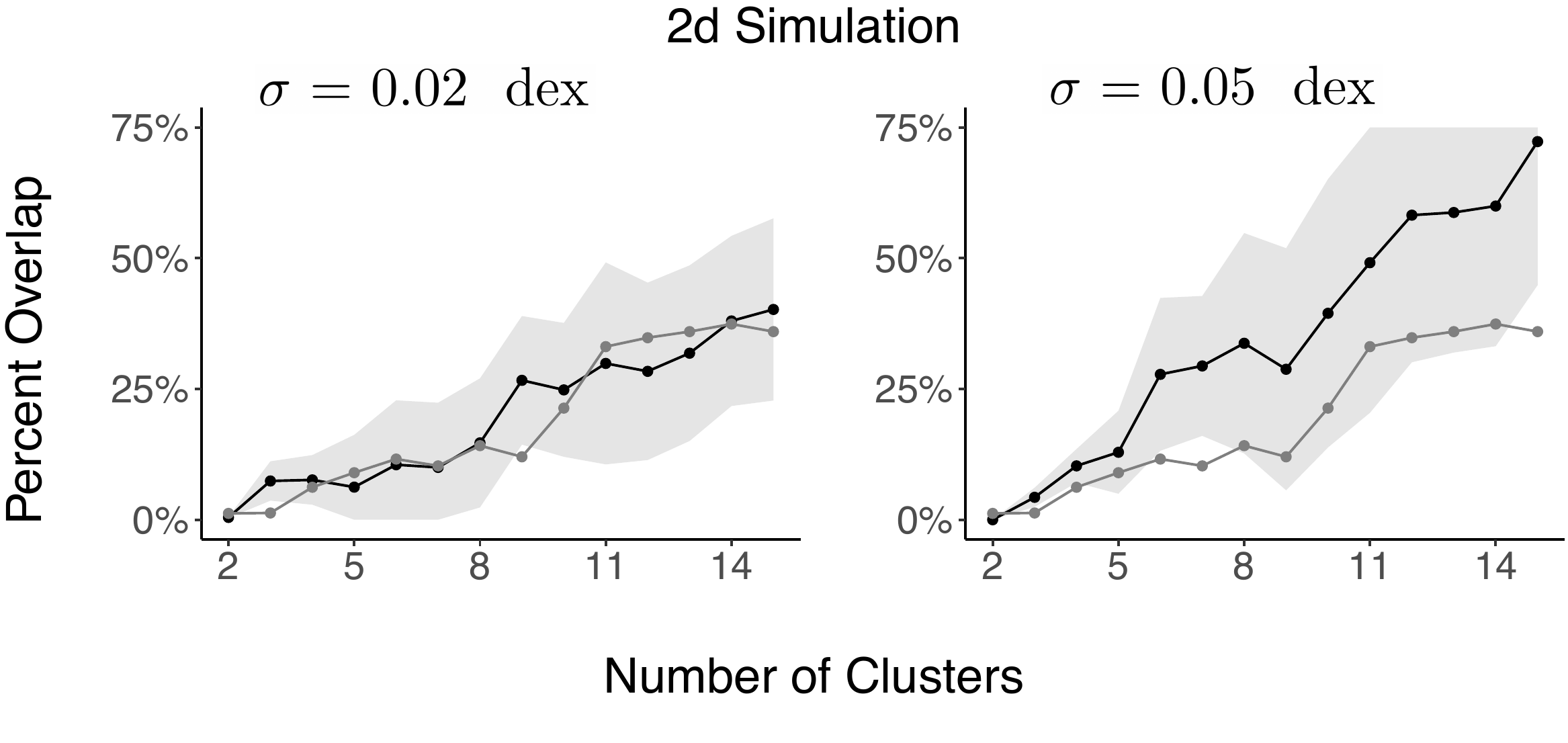}\\
     \includegraphics[width=.86\textwidth]{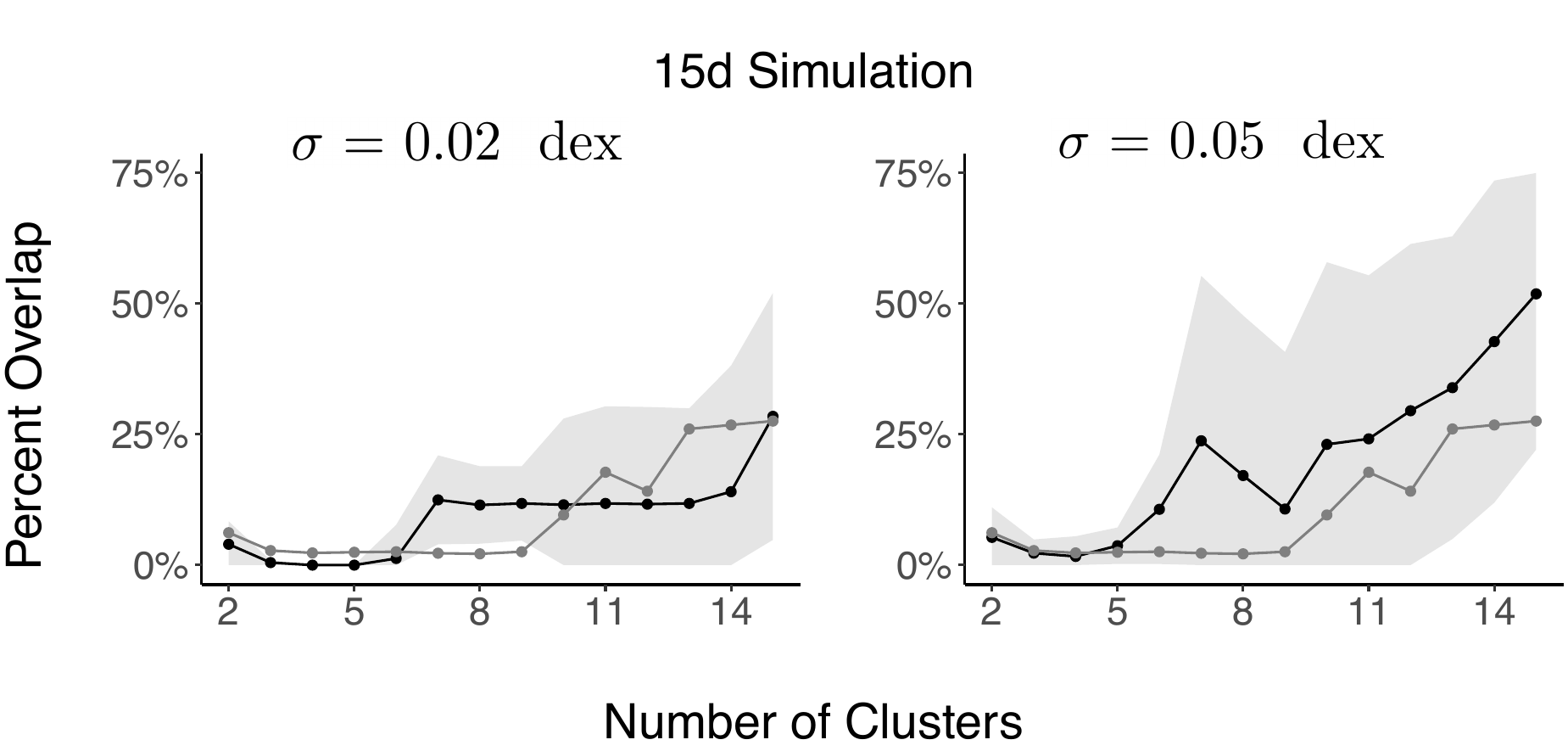}
\caption{The black points and line correspond to the median percent that clusters defined in an error-convolved (\textbf{top}) 2-dimensional and (\textbf{bottom}) 15-dimensional abundance space overlap in the age--\rbirth\ plane. For each star, new abundance measurements are drawn from a normal distribution with the true abundance value as the mean and a standard deviation of (\textbf{left}) 0.02 dex and (\textbf{right}) 0.05 dex. The grey ribbon captures one standard deviation about the median percent that clusters overlap in the age--\rbirth\ plane. When simulating errors of 0.02 dex, clusters still stay separated in the age--\rbirth\ plane and the separation is comparable to that given by clusters found with no abundance error (grey points and line). Adding an error of 0.05 dex in each dimension affects our ability of finding separate birth information slightly more in the 2-dimensional simulation than the 15-dimensional simulation.}
\label{fig:error}
\end{figure*}

Section \ref{sec:results_clustering} showed that under certain formation conditions and no observational limitations, that abundances trace stellar birth information \rbirth\ and age. In this section we examine some of the consequences of observational limitations. 

\subsection{Incorporating Measurement Uncertainty}

Current day element abundance measurements are reported with uncertainties of $\approx$ 0.02-0.05 dex \citep{apogeedr16,2020AJ....160..120J}. Consequently, we examine how the clustering changes once we incorporate errors in the chemical abundances and how this impacts our ability to trace birth properties with observational data.

For each star in both the 2d and 15d simulations, we redraw a new set of element abundances, each from a Gaussian distribution where the standard deviation is representative of the measurement uncertainty. We test two precision regimes: $\sigma_{\mbox{err}}$ = 0.02 and 0.05 dex. The left column of Figure \ref{fig:error} shows the overlap in the age--\rbirth\ plane of clusters defined in the 2-dimensional (top) and 15-dimensional (bottom) abundance simulations when the abundances are modified with an equivalent of 0.02 dex error in each abundance direction (black points and line). With the current best observational error of $\sigma_{\mbox{err}}$ = 0.02 dex, the clusters defined in both the 2-dimensional and 15-dimensional abundance space retain separate birth properties, similar to the overlap found when the simulations have no error added (grey points and line). When abundances are redrawn with  observational errors of $\sigma_{\mbox{err}}$ = 0.05 dex for each data point (right column), we find that the majority of clusters from the higher dimensional simulation retain more separate birth information compared to those found in the 2-dimensional simulation.

\subsection{Modifying the sample size} \label{sec:results_suberrors}
\begin{figure*}
     \centering
     \includegraphics[width=1\textwidth]{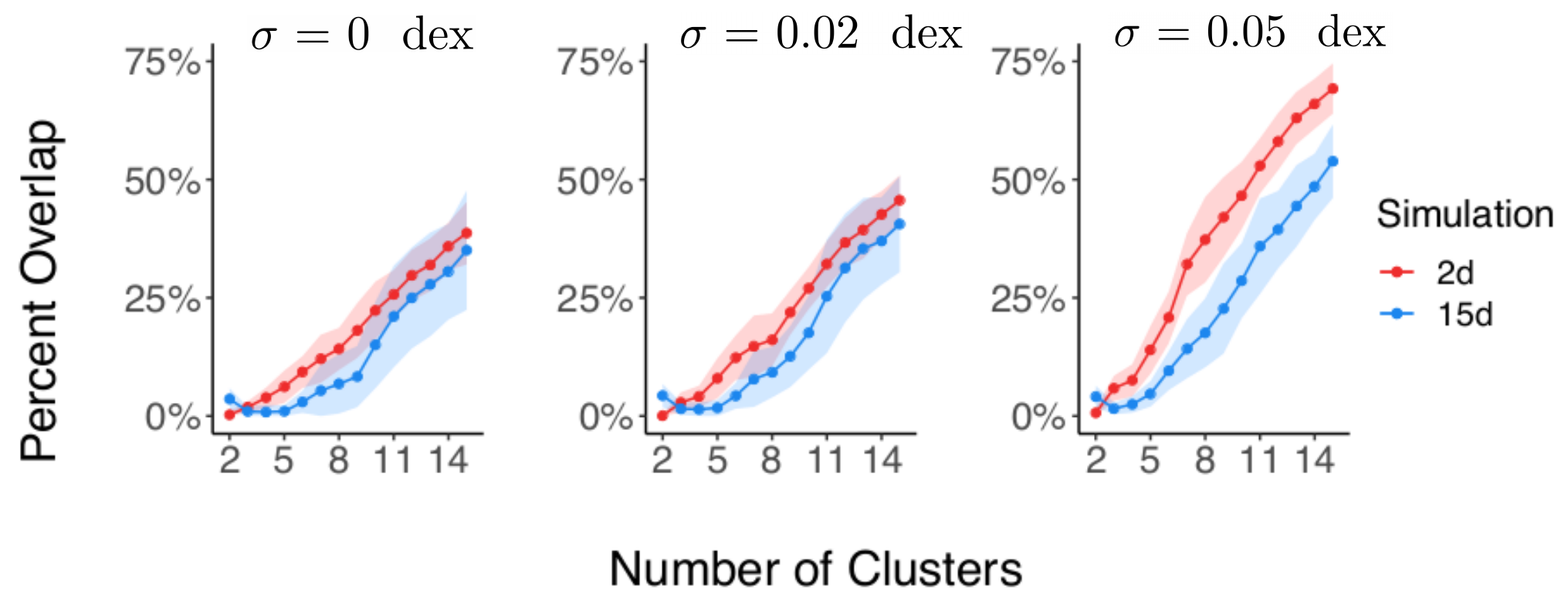}
\caption{The mean percent of 50 subsampling replications that clusters defined in a 30,000 stellar sample of an error-convolved 2-dimensional (red) and 15-dimensional (blue) abundance space overlap in the age--\rbirth\ plane. The errors added are equivalent to the observational best case scenario (\textbf{middle}, 0.02 dex) and average observational error (\textbf{right}, 0.05 dex) in each dimension. \textbf{Left} has no error added. The ribbon shows one standard deviation in percent overlap for the 50 monte carlo simulations.}
\label{fig:subError}
\end{figure*}

While current large surveys have captured many millions of stars, we have so far examined only about 30,000 stars with precise abundance measurements, within a narrow evolutionary state \citep[e.g. \apogee\ DR14 RC catalog][]{bovy2014apogee}. To test more generally how useful chemical abundances are for linking to birth properties, we need to examine the impact of sample size.

So far in this work, we have been working with $\sim$229,000 and $\sim$44,000 star particles for the 2d and 15d simulations respectively. Now we examine how the clusters defined in abundance space --- both with and without the addition of errors --- change in the age--\rbirth\ plane for a random sub-sample of 30,000 stars throughout the whole disk. Figure \ref{fig:subError} shows the percent overlap for both the 2-dimensional (red) and 15-dimensional (blue) subsampled simulations with no error (left), the equivalent of 0.02 dex error (middle) and 0.05 dex error (right) in each dimension. In order to test the consistency of these results, we subsample with 50 replications. The mean and standard deviation of the median percent overlap are shown as a point and ribbon.

We can see from Figure \ref{fig:subError} that as error increases, the amount the 2-dimensional clusters overlap in the age--\rbirth\ plane also increases. This indicates that for the sample size and error used in paper I of this series \citep{2020Ratcliffe}, only two abundance dimensions are not enough to recover separate birth time and place groups observationally for more than $\approx$ 6 clusters. 

On the other hand, the clusters in the 15-dimensional subsample are less affected by error than the 2-dimensional simulation. We can see that as error increases, the amount each cluster overlaps with other clusters in the age--\rbirth\ plane stays more consistent than the clusters in the 2-dimensional subsampled simulation. While the clusters from both simulations trace less birth information in the presence of subsampling, we see that the recovery of separate birth place and time clusters is still possible with the inclusion of more abundances and nucleosynthetic families, especially for a smaller number ($\leq\,\approx$10) of clusters.
\begin{figure*}
     \centering
     \includegraphics[width=1\textwidth]{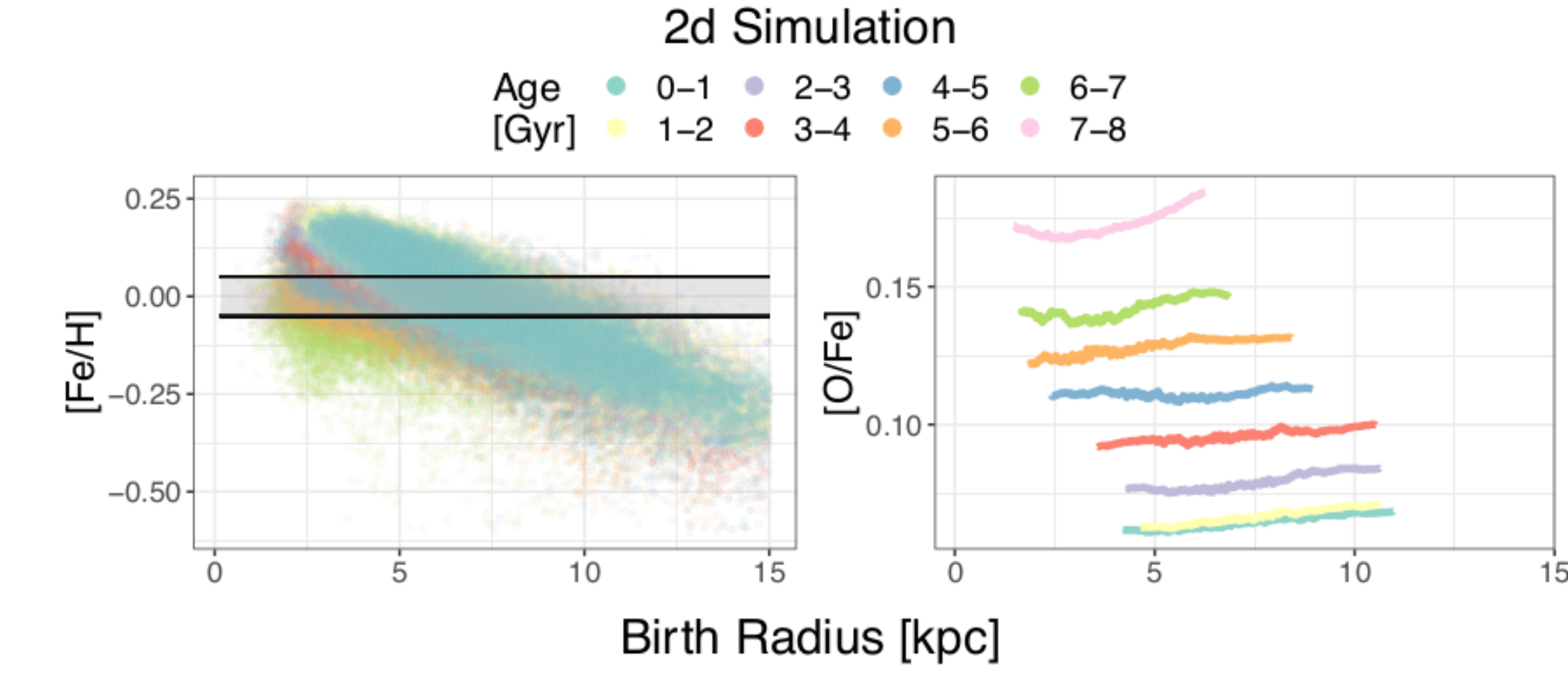}
\caption{(\textbf{Left}) The \feh--\rbirth\ plane colored by age for the 2-dimensional simulation. The black lines and grey area mark off the solar metallicity stars, which we consider to be $\pm$0.05 dex in \feh. (\textbf{Right}) the running mean of [O/Fe] of the solar metallicity stars across \rbirth\ colored by age selected from within the horizontal lines at left. We see that for a given bin of metallicity, stars of different ages separate out and form either approximately quadratic or linear relations.}
\label{fig:Rbirthxfe_2d}
\end{figure*}

\section{Results III: Modeling how observable stellar properties relate to birth properties} \label{sec:results_regression}

\begin{figure*}
     \centering
     \includegraphics[width=0.92\textwidth]{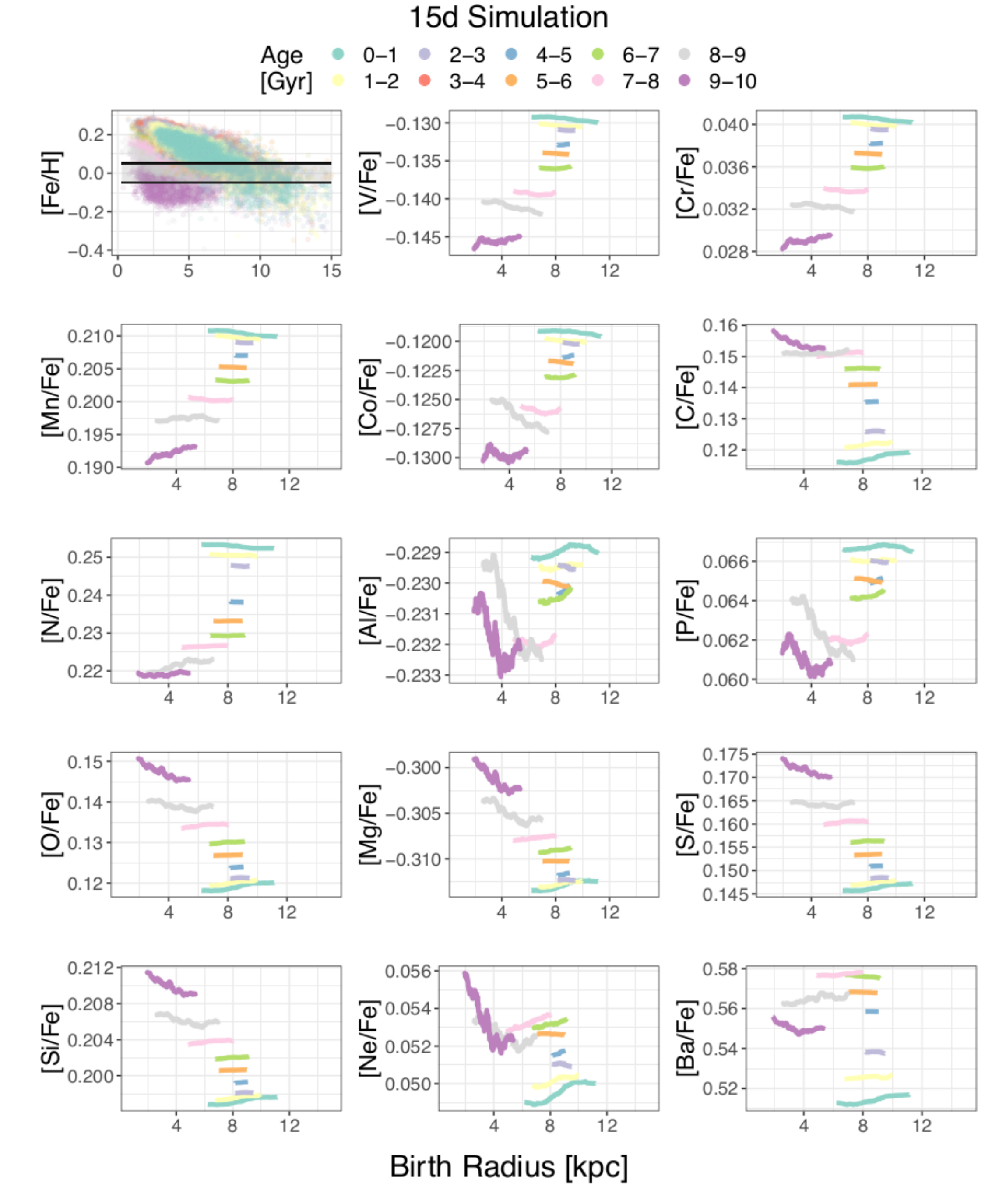}
\caption{(\textbf{Top left})  The \feh--\rbirth\ plane colored by age for the 15-dimensional simulation. The black lines and grey area mark off the solar metallicity stars, which we consider to be $\pm$0.05 dex in \feh. All other plots show the running mean of \xfe\ of the solar metallicity stars accross \rbirth\ colored by age. Similar to the 2-dimensional simulation shown in Figure \ref{fig:Rbirthxfe_2d}, solar metallicity stars of different ages separate into approximately quadratic or linear curves.}
\label{fig:Rbirthxfe_hd}
\end{figure*}

In the previous sections, we showed that clusters in abundance space correspond to discrete groups in birth time and space. Now we quantify how chemical abundances relate to age and \rbirth.

The left panel of Figure \ref{fig:Rbirthxfe_2d} shows the 2-dimensional simulation in the \feh--\rbirth\ plane, with each point colored by the star's age. In order to visually examine if an age--abundance--\rbirth\ relationship exists, we examine running \xfe\ means of solar metallicity stars (which are taken to be stars within the grey band) as a function of \rbirth, divided into age bins. The right panel of Figure \ref{fig:Rbirthxfe_2d} shows that for the 2-dimensional simulation, there is a strong age--\alphafe\ relation in solar metallicity stars that is approximately quadratic for older stars and linear for younger stars. Thus, given a fixed \feh, we anticipate that ages can be determined from abundances. Similarly for the 15-dimensional simulation shown in Figure \ref{fig:Rbirthxfe_hd}, each age group has its own unique polynomial trend in [X/Fe]--\rbirth.

\begin{figure*}
     \centering
     \includegraphics[width=.9\textwidth]{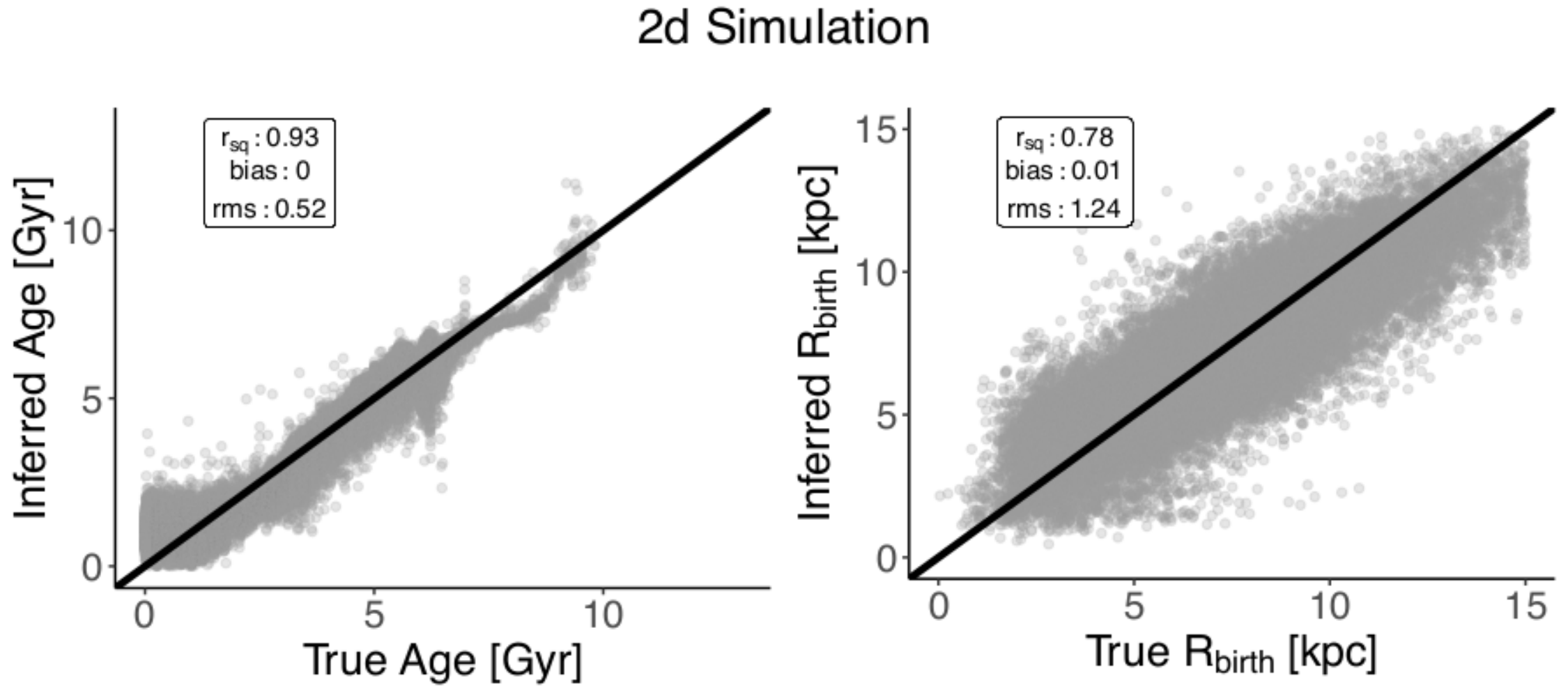}\\
     \includegraphics[width=.9\textwidth]{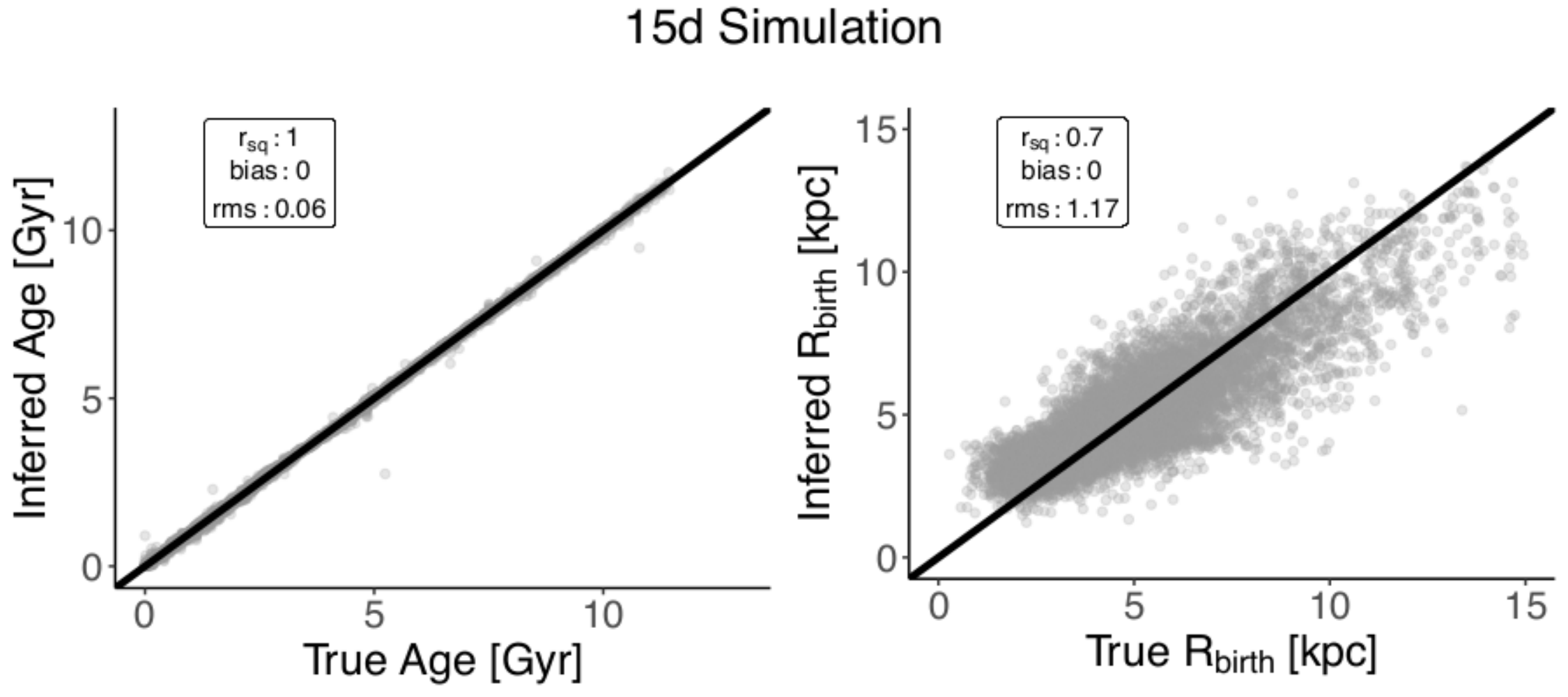}
\caption{(\textbf{Left}) The inferred age of the (\textbf{top}) 2-dimensional and (\textbf{bottom}) 15-dimensional simulation using a second order polynomial in (\feh, \xfe) plotted against the true age of the star. With the additional abundance information provided in the 15d simulation, we are able to accurately and precisely estimate age, showing that abundances are chemical clocks. (\textbf{Right}) Inferred \rbirth\ of the (\textbf{top}) 2-dimensional and (\textbf{bottom}) 15-dimensional simulation using a second order polynomial in (\feh, age) plotted against the true \rbirth\ of the stars. With just \feh\ and age, we can infer a star's \rbirth\ to within just over 1 kpc.}
\label{fig:regression}
\end{figure*}

This visual analysis done in Figures \ref{fig:Rbirthxfe_2d} and \ref{fig:Rbirthxfe_hd} leads to the conclusion that ages can be determined from abundances alone. To quantify this relationship, we use a simple second order polynomial to estimate age from ([X/Fe], [Fe/H]). The model for the 2-dimensional simulation is
\begin{align*}
    age = & a_1 \feh^2 + a_2 \alphafe^2 + \\
          & a_3\feh\times\alphafe\ + a_4 \feh +\\
          & a_5 \alphafe + a_0,
\end{align*}

where the $a_i$s are the coefficients. The model for the 15-dimensional simulation is similar, however with the inclusion of more abundances. The left column of Figure \ref{fig:regression} shows the inferred age from the polynomial regression versus the true age of stars in the simulation. Even with just two abundances (top row), we are able to estimate age within $\pm$0.52 Gyr. With the addition of more abundance information (bottom row), we find that we are able to accurately estimate age from 15 abundances to within $\pm$0.06 Gyr. 

\begin{figure*}
     \centering
     \includegraphics[width=0.98\textwidth]{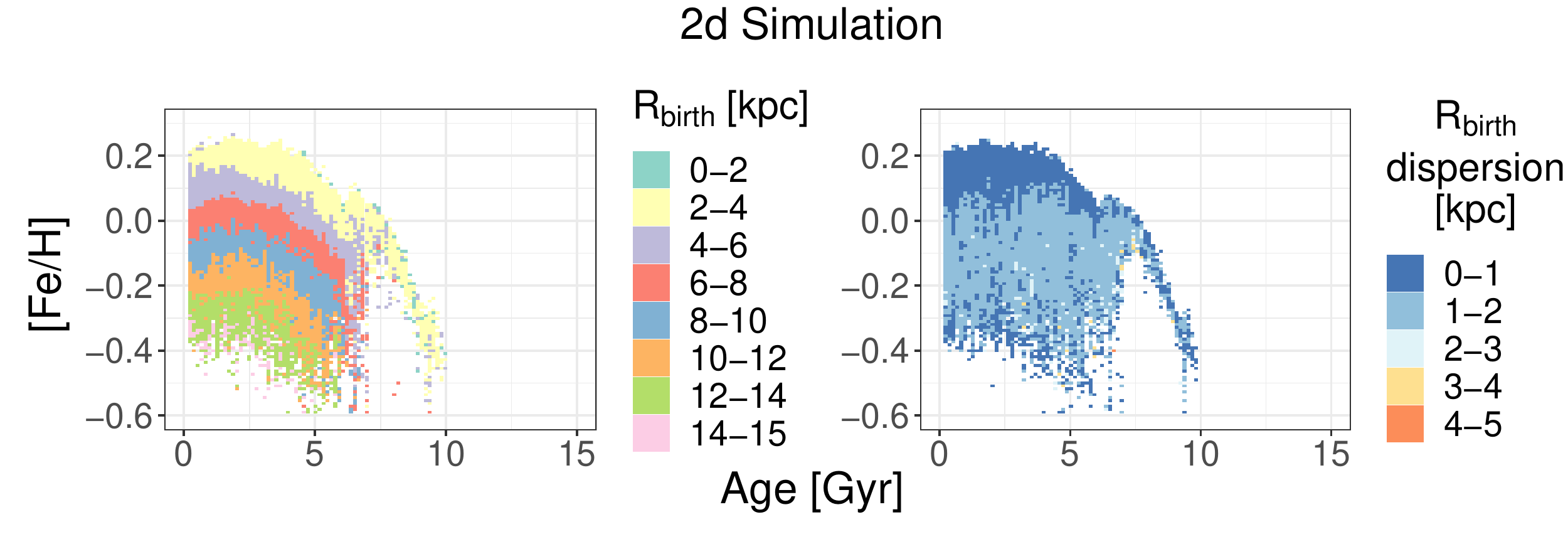}\\
     \includegraphics[width=0.98\textwidth]{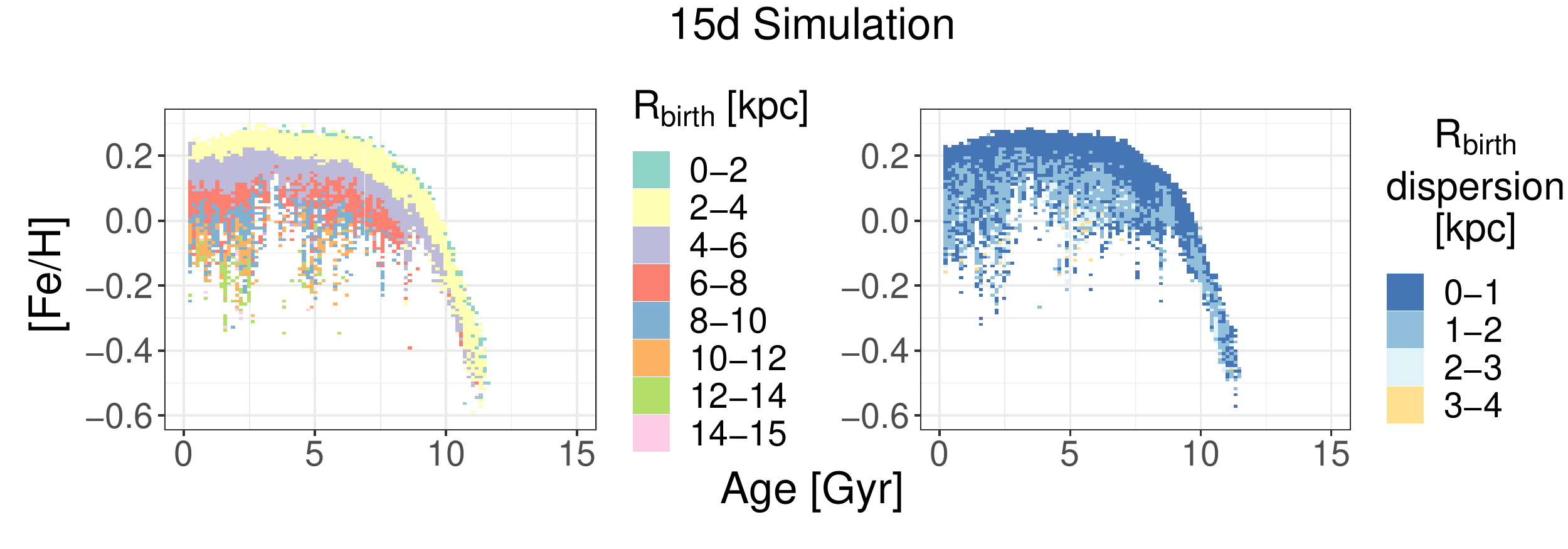}\\
\caption{The \feh--age plane colored by (\textbf{left}) \rbirth\ and (\textbf{right}) \rbirth\ dispersion for the (\textbf{top}) 2- and (\textbf{bottom}) 15-dimensional simulations. The low dispersion in this plane indicates that \feh\ and age alone can determine \rbirth\ accurately.}
\label{fig:RbirthDisps}
\end{figure*}

We also wish to test how well we can quantify the relationship between age, \rbirth, and abundances. Given the low \rbirth\ dispersion in the \feh--age plane shown in Figure \ref{fig:RbirthDisps}, we use a second order polynomial model to estimate \rbirth\ from (\feh, age). The model for both simulations is
\begin{align*}
    r_\text{birth} = & a_1 \feh^2 + a_2 age^2 + a_3 \feh\times age +\\
          & a_4 age + a_5 \feh + a_0,
\end{align*}
where the $a_i$s are the coefficients.
The right panels of Figure \ref{fig:regression} reveals that we can predict \rbirth\ to within $\pm$1.24 kpc for the 2-dimensional simulation (top) and within $\pm$1.17 kpc for the 15-dimensional simulation (bottom). The inclusion of the additional abundances increases the accuracy by 0.06 kpc for the 15-dimensional simulation and only 0.01 kpc for the 2-dimensional simulation. We again see that additional abundance information helps inform more about stellar birth properties, however the difference is not as drastic as it was in estimating age.

While we do not fit for the best model, we believe that this simple second order polynomial relationship between age, abundances, and \rbirth\ cannot be drastically improved upon. For a given value of \feh, \xfe, and age, we find that the intrinsic dispersion in \rbirth\ is $\approx$ 1.1 kpc for the 15d simulation and $\approx$ 1.2 kpc for the 2d simulation. Thus, ages and abundances alone will not be able to estimate \rbirth\ more accurately. This could be due to asymmetries causing abundance distributions to not lie in perfect annuli about the galactic center, reducing the tightness of the relationship between \rbirth\ and abundances.

With the inclusion of 0.05 dex abundance error, we find that our age estimates decrease in accuracy to about $\pm 0.76$ Gyr for the 15d simulation. With the addition of 0.05 dex abundance error and age error of 30\%, the \rbirth\ accuracy of the 15d simulation decreases to $\pm 1.31$ kpc. This shows that according to our model, estimating \rbirth\ from abundances and age is less sensitive to noise than estimating age from abundances.

\section{Discussion --- Implications for future applications to the Milky Way}\label{sec:discussion}

Paper I of this series \citep{2020Ratcliffe} used hierarchical clustering in the 19-dimensional abundance space of 30,000 red clump stars in the Milky Way catalogued by \apogee\ DR14. In that work, we found that up to six clusters have statistically significant different mean ages and distinct spatial distributions. The goal of this work is aimed to interpret those results and determine if clusters observed in chemical space correspond to physically meaningful groups. With the use of simulations, we are able to test the potential and current ability of linking current stellar properties (\feh, \xfe) to their birth properties (\rbirth, age). 

\subsection{Empirical Context}

The simulations used in this work (2-dimensional \buckSim\ and 15-dimensional \buckHD) are Milky Way analogs from the NIHAO-UHD suite. Both simulations were bulge-dominated systems up to redshift $z\geq 1$ with prominent stellar disks forming 7-8 Gyr ago. The formation of the $\alpha-$sequences was due to a gas-rich merger, with the high-$\alpha$ sequence forming during the early galaxy and the low-$\alpha$ sequence forming after the merger. The main differences between these simulations are a slightly different formation history (sampling valid formation histories of Milky Way like galaxies) as well as an updated chemical enrichment prescription for the 15d model galaxy. The modification made to chemical enrichment prescriptions are described in \citet{2021BuckHD_chemEnrich} and enabled us to follow 15 different elements while at the same time leaving global galaxy properties such as star formation history, stellar mass, and disk size unaffected. We believe that these simulations are representative of the Milky Way due to their formation history and age gradient in the abundance plane.

Even though simulations provide us with particles and not individual stars, they allow us to examine the relationship between observable chemical properties, age, and birth location. Thus we focus on disk particles in our work.

\subsection{Clustering Approaches}\label{sec:discussion_clustering}

In this work, we focused on three ways of grouping stars --- hierarchical clustering, EnLink, and binning in the \alphafe--\feh\ plane. Section \ref{sec:results_whyCluster} shows that binning in just (\feh, \alphafe) does not effectively link abundance information to birth properties, especially when there is higher dimensional abundance data available. 

Of the clustering methods we explore, hierarchical clustering is advantageous for observational work. Section \ref{sec:enlink_2d} shows that leveraging density with an adaptive distance metric in the 2-dimensional simulation is the best way for chemical clusters to correspond to distinct groups in the age--\rbirth\ plane. However, due to the curse of dimensionality, leveraging density in a 15-dimensional space is difficult and unrealistic. We attempted to run EnLink in the 15-dimensional chemical abundance space, however clustering results were inconclusive and did not define many stars to a cluster. Additionally, for EnLink (or any other density based clustering method) to be used correctly on Milky Way catalogs, the survey selection function would need to be accounted for, as the selection function would possibly alter the distribution of stars in abundance space. Furthermore, we found that EnLink performed poorly when the density structure in the 2-dimensional abundance plane vanished after the addition of observational uncertainty. 

\subsection{Likelihood of success --- Comparison to other simulations}\label{sec:victor}

\begin{figure*}
     \centering
     \includegraphics[width=\textwidth]{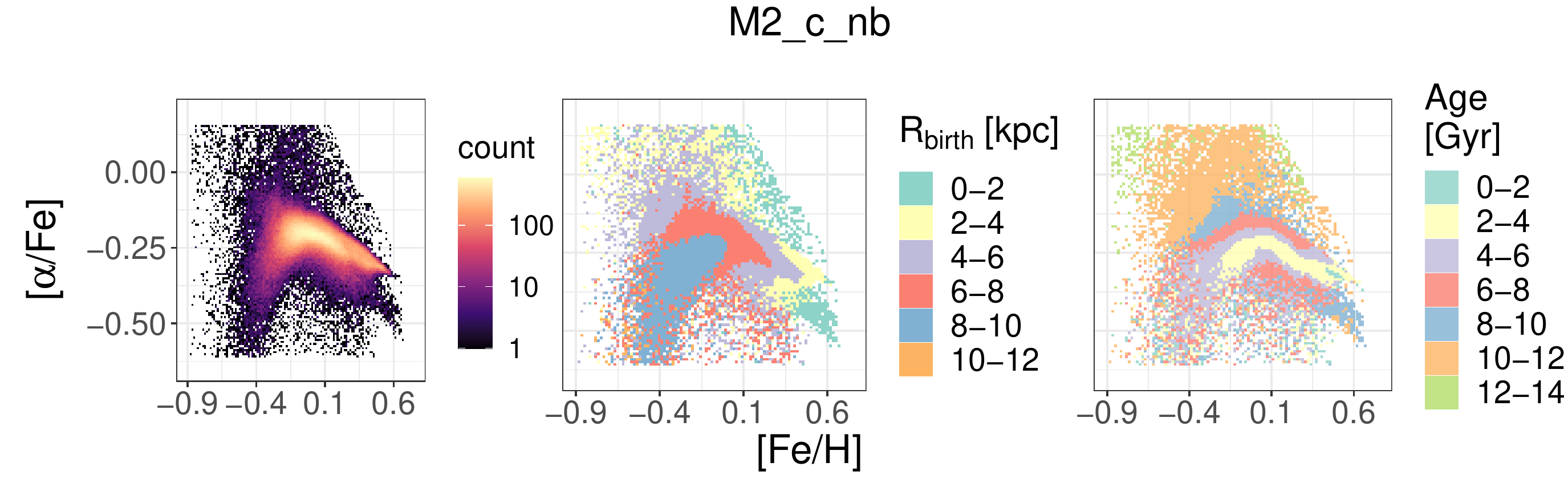}\\
\caption{The \alphafe\--\feh\ plane colored by (\textbf{left}) density, (\textbf{middle}) \rbirth, (\textbf{right}) age for the \cite{2021Silva_simulation} simulation using clumpy star formation. The formation history produces quadratic age and \rbirth\ distributions in the abundance plane, and causes the simple relationship between age, \rbirth, and clustered abundances to disappear.}
\label{fig:victor}
\end{figure*}

In this paper, we have focused on the relationship between abundances and birth properties of stars when the chemical bimodality is caused by a merger and successive dilution of the interstellar medium. We also consider the relationship when a galaxy is formed by clumpy star formation\footnote{Note, the galaxies simulated within the NIHAO project also go through a clumpy phase \citep{Buck2017} in agreement with observed clumpy galaxies at high redshift. However, for the NIHAO feedback scheme those clumps are agglomerations of young stars and only appear in stellar light not in stellar mass.}. We examine the $N$-body+Smooth Particle Hydrodynamics simulation of the formation of an isolated galaxy outlined in \cite{2021Silva_simulation}. Star-forming clumps at high redshift start forming low-$\alpha$ stars, then quickly self-enrich in $\alpha$-elements due to their high star formation rate density and produce the high-$\alpha$ sequence while the low-$\alpha$ sequence is produced by radially distributed star formation. After about 4 Gyr, the clumps become less efficient, and the high-$\alpha$ sequence stops growing \citep{Clarke2019}. For more detailed information of the simulation, see \cite{2021Silva_simulation} and \cite{2021MNRAS.503.1418F}.

Figure \ref{fig:victor} shows the \alphafe--\feh\ plane for a 230,000 particle subsample of simulation M2\_c\_nb, which undergoes clumpy star formation. Similar to the 2- and 15-dimensional simulations, there is a linear trend between the abundances and \rbirth, where \rbirth\ decreases as \feh\ increases. Age, however, does not appear to have a simple relationship between the two abundances. For instance, age decreases as \alphafe\ decreases for solar metallicity stars at higher values of \alphafe, but the relationship is reversed for lower \alphafe.

We find that in this simulation, age and \feh\ are able to predict \rbirth\ within $\pm0.72$ kpc --- about 40\% better than the precision for the 2- and 15-dimensional simulations focused on in this work. Again, we also see that the addition of other abundances does not notably change our ability to estimate \rbirth, where the accuracy only increases by 0.01 kpc when \alphafe\ is included in the regression. This shows that the formation history in all three simulations sets an underlying relationship with age, \feh, and \rbirth, where if we know the metallicity and age of a star, we can determine where it was born. 

However, while [Fe/H] and age are a link to \rbirth, in this particular simulation the star formation history gives rise to a more complex relationship between the abundances ([Fe/H], \alphafe) and age (Figure \ref{fig:victor}). Therefore chemically similar groups of stars identified using hierarchical clustering no longer correspond to separate groups in the age--\rbirth\ plane in this scenario.

\begin{figure*}
     \centering
     \includegraphics[width=.7\textwidth]{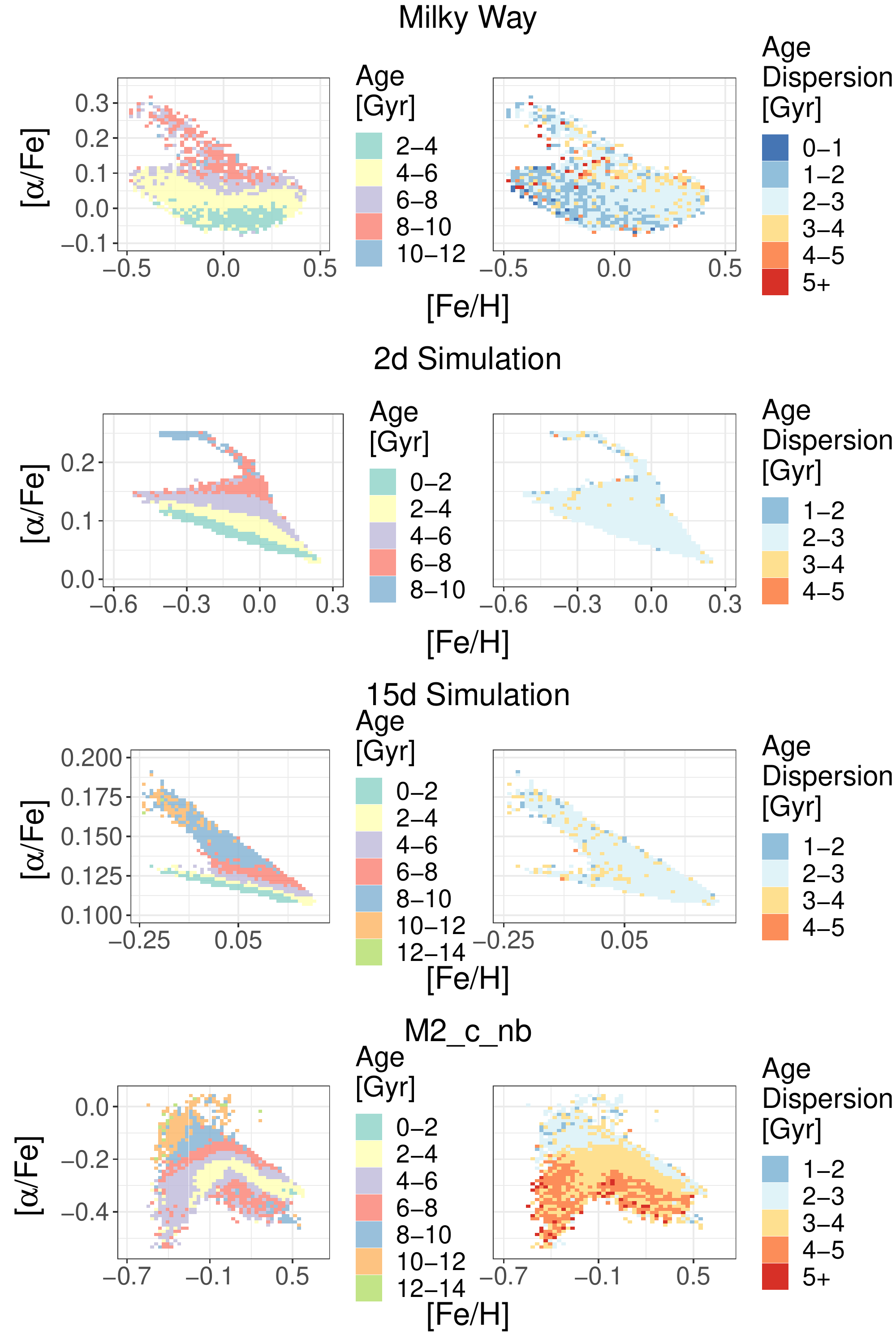}
\caption{(\textbf{Top}) The Milky Way in the \alphafe--\feh\ plane using \cite{yuxi2021universal} ages and abundances. The right panel shows the standard deviation of ages within each bin. The ages for the (\textbf{second from top}) 2d simulation,  (\textbf{third from top}) 15d simulation and (\textbf{bottom}) M2\_c\_nb are redrawn from a normal distribution with their true age as the mean and a standard deviation equivalent to the median uncertainty of low-$\alpha$ stars from the \cite{yuxi2021universal} catalog. Only bins containing at least ten stars are shown in each plot above.}
\label{fig:ageDisp}
\end{figure*}

In order to determine the ability to extend our conclusions to the Milky Way, we compare the Milky Way's age and age dispersion in the \alphafe--\feh\ plane to the three simulations with simulated ``observational" ages by redrawing from a normal distribution with their true age as the mean and a standard deviation of $\approx 2.6$ Gyr --- the median uncertainty of low-$\alpha$ stars from the \cite{yuxi2021universal} catalog. When simulating observational ages for the 2- and 15-dimensional simulations, the dispersion in age across \alphafe--\feh\ is uniformly low (see middle rows of Figure \ref{fig:ageDisp}), while M2\_c\_nb has an increase in age dispersion as \alphafe\ decreases (see bottom row of Figure \ref{fig:ageDisp}). The Milky Way (shown in the top row of the same figure using ages and abundances from \cite{yuxi2021universal}) has a consistently small dispersion in age of about 2-3 Gyr, with the dispersion being slightly smaller for low (\feh, \alphafe) --- the reverse of M2\_c\_nb's dispersion trends. On the other hand, similar to the 2d and 15d simulations focused on in this paper, the age of the stars in the Milky Way increases as \alphafe\ increases for a given value of \feh.

The question of which simulation most closely matches the Milky Way’s star formation history still requires further investigation. The selection cuts described in Section \ref{sec:selection_cuts} produce different density trends in the \alphafe--\feh\ plane for the three simulations and observational data, where some samples have both high- and low-$\alpha$ stars (e.g. the 15-dimensional simulation), while others primarily consist of stars with lower values of \alphafe\ (e.g. M2\_c\_nb). There is room for exploration into how the Milky Way compares to the different \alphafe--\feh\ trends each simulation produces with different selection cuts, however, we find that our results are consistent under different cuts to capture disk stars.

This exploration shows that resolving distinct birth properties of chemically similar stars requires a small dispersion in age and \rbirth\ trends in the \alphafe--\feh\ plane. Since the Milky Way has been shown to have age trends in the \alphafe--\feh\ abundance plane with small dispersion, we believe the conclusions we have drawn in this paper are relevant to the Milky Way. We conclude that the six clusters found in our previous work \citep{2020Ratcliffe} using hierarchical clustering in the 19-dimensional \apogee\ red clump sample form discrete groups in birth time and place.

\subsection{Limitations and future work}

There are some limitations to our approach. In Section \ref{sec:results_suberrors} we discuss the effect of subsampling data with observational errors. However, we do not take into account the complexity of survey selection functions. Additionally, in Section \ref{sec:results_errors} we explore how errors affect our clustering results by adding 0.02 dex or 0.05 dex error to every abundance. Realistically though, some abundances are measured more accurately than others. 

Section \ref{sec:discussion_clustering} discussed how density based clustering failed when structure vanished due to measurement error or in the high dimensional abundance space. For future work, combining hierarchical clustering with an adaptive distance metric would be interesting to explore. Partnering an adaptive distance metric with hierarchical clustering would avoid the problems of estimating density and determining the best distance metric in a high dimensional abundance space, and could potentially provide even more striking results. 

As simulation resolution continues to increase, in the future it would also be useful to explore if satellite debris could be picked up by abundance clustering and complement clustering analysis done in action space \citep[such as in ][]{2021arXiv210408185W}. In the simulations we use in this study, only $\approx$ 20 stellar particles have \rbirth $\geq$ 20 kpc. With future datasets and simulations in mind, testing to see if accreted material differ in abundance space could be useful to determine accreted debris in the Milky Way. 

\section{Summary and Conclusions}\label{sec:conclusion}
Our main results are summarized below: 

\begin{itemize}
    \item We find with just \feh\ and \alphafe\ alone we can trace separate \rbirth--age clusters, with the separation being more distinct when we include more abundances, as demonstrated by our 15 element simulation \buckHD\ where we find nearly completely separate groups for up to $\approx$ 10 clusters (Figure \ref{fig:contour_3panel}). Considering current day observational uncertainty and sampling constraints, higher dimensional abundance data is necessary to trace birth properties from chemical abundance data. Clusters from the subsampled 2-dimensional abundance simulation with 0.05 dex error had substantial overlap with other clusters in the age--\rbirth\ plane. On the other hand, clusters found in the subsampled 15d simulation with 0.05 dex error had almost no overlap in age and \rbirth\ when finding six or fewer clusters (Section \ref{sec:results_errors}; Figure \ref{fig:subError}). 
    
    \item The clusters found in this paper presumably trace not only separate areas in the age--\rbirth\ plane, but also link to different underlying physical properties. Stars separated by hierarchical clustering preserved a clear metallicity gradient as a function of age for a given \rbirth, whereas groups defined by binning in the \alphafe--\feh\ plane lost the \feh\ gradient and increased their age dispersion (Figure \ref{fig:meanRbirthVage_feh}).
    
    \item Chemical clusters of high- and low-$\alpha$ stars separate differently in the age--\rbirth\ plane. Clusters defined with low-$\alpha$ stars separate both as a function of age and \rbirth, showing low-$\alpha$ stars are born throughout the galaxy at different radii. High-$\alpha$ stars are older ($> 7$ Gyr) and are born near the galactic center, but separate as a function of age (see Figures \ref{fig:contour_3panel} and \ref{fig:enlink_clusters}). 
    
    \item Using a simple second order polynomial regression, we quantify the relationship between observable abundance labels and birth property outputs (Section \ref{sec:results_regression}). We model age as a function of (\feh, \xfe), and can infer a star's age to a precision of $\pm 0.52$ Gyr for the 2-dimensional abundance simulation and $\pm 0.06$ Gyr for the 15-dimensional abundance simulation. We also model \rbirth\ as a function of (\feh, age), and infer it to a precision of $\pm 1.24$ kpc and $\pm 1.17$ kpc for the 2- and 15-dimensional abundance simulations respectively. 
    
    \item The ability to reconstruct stellar groups born in different times and places from their abundances is determined by the formation history of the galaxy. When formation conditions lead to age and \rbirth\ trends in the abundance plane with small dispersion, we find that there is a simple connection between clustered abundances and discrete birth times and places. Under clumpy star formation however, the simple relationship vanishes (Section \ref{sec:victor}). 
    
    \item Our comparison of three simulations implies that the low dispersion of age  across the \alphafe-\feh\ plane of the Milky Way indicates that the Milky Way's star formation history is sufficiently quiet and that clustering in abundance will correspond to birth associations in time and location (Figure \ref{fig:ageDisp}).
    
\end{itemize}

We seek to examine how abundance structure links to birth properties. We find that there is a simple relationship between age and chemical abundances, which agrees with previous work \citep[e.g.][]{Ness2019}. \rbirth\ cannot be tested as we can do for age --- we never have direct access to this quantity in observations. From our regression however, we see age and ([Fe/H], \xfe) link us to \rbirth\ in the simulations. Indeed this analytical formalism has been adopted in models of radial migration \citep[e.g.][]{Frankel2018,2019minchev}. We examine the \rbirth--age distribution further using the idea of abundance clustering, which we seek to see if it links to underlying physical processes. 

This work highlights how we might use clustering of high dimensional abundance measurements in large surveys to infer groups of different birth place and time, and the impact of measurement uncertainty in working with the observational data. 

\section{Acknowledgements} 
Melissa K Ness acknowledges support from the Sloan Foundation Fellowship. 
Tobias Buck acknowledges support from the European Research Council under ERC-CoG grant CRAGSMAN-646955. This research made use of {\sc{pynbody}} \citet{pynbody}.
We gratefully acknowledge the Gauss Centre for Supercomputing e.V. (www.gauss-centre.eu) for funding this project by providing computing time on the GCS Supercomputer SuperMUC at Leibniz Supercomputing Centre (www.lrz.de).
This research was carried out on the High Performance Computing resources at New York University Abu Dhabi. We greatly appreciate the contributions of all these computing allocations.
K.V.J. is supported by NSF grant AST-1715582.
B.S. is supported by NSF grant DMS-2015376.
VPD and LBS are supported by STFC Consolidated grant \#ST/R000786/1

\section{Appendix - additional figures}
Here we include abundance--age plots colored by \rbirth\ for both the 2d and 15d simulations. These plots are similar to Figures \ref{fig:Rbirthxfe_2d} and \ref{fig:Rbirthxfe_hd}, however the coloring and x-axis are switched. 

\begin{figure*}
     \centering
     \includegraphics[width=\textwidth]{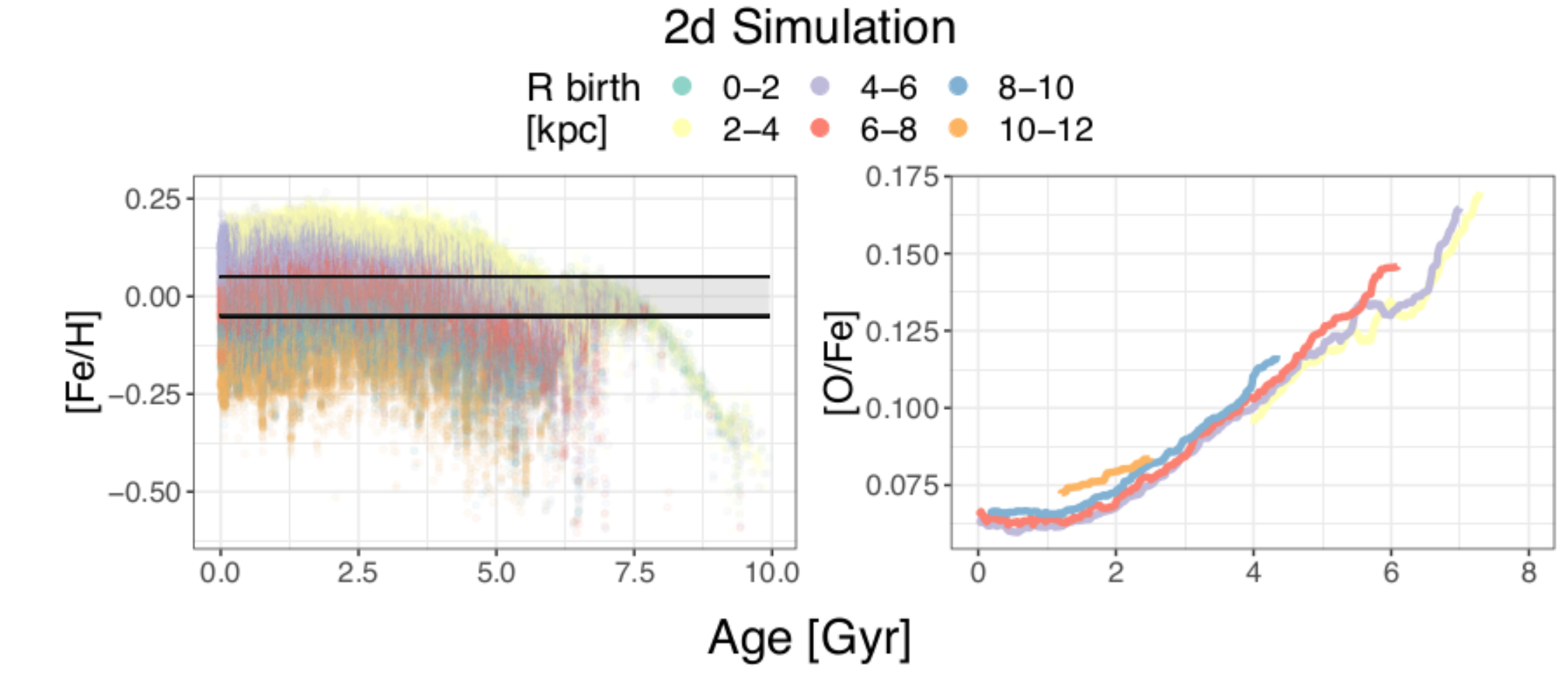}
\caption{(\textbf{Left}) The \feh--age plane colored by \rbirth\ for the 2-dimensional simulation. The black lines and grey area mark off the solar metallicity stars, which we consider to be $\pm$0.05 dex in \feh. (\textbf{Right}) the running mean of [O/Fe] of the solar metallicity stars across age colored by \rbirth\ selected from within the horizontal lines at left. For a given bin of metallicity, stars clearly have a polynomial trend in \xfe--age.}
\label{fig:Agexfe}
\end{figure*}

\begin{figure*}
     \centering
     \includegraphics[width=0.92\textwidth]{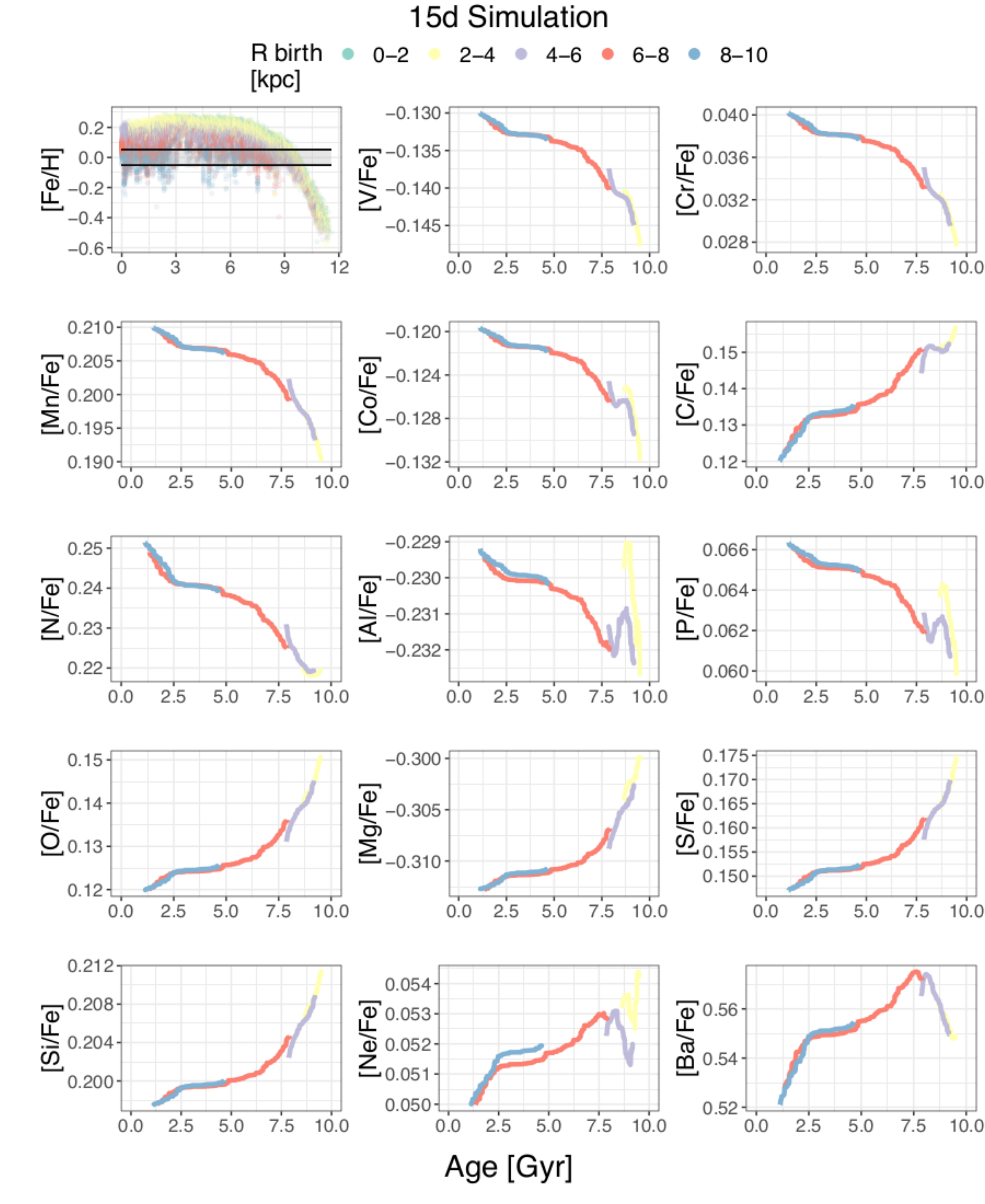}
\caption{(\textbf{Top left})  The \feh--age plane colored by \rbirth\ for the 15-dimensional simulation. The black lines and grey area mark off the solar metallicity stars, which we consider to be $\pm$0.05 dex in \feh. All other plots show the running mean of \xfe\ of the solar metallicity stars accross age colored by \rbirth. Similar to the 2-dimensional simulation shown in Figure \ref{fig:Agexfe}, solar metallicity stars of different ages separate into different polynomial curves.}
\label{fig:Agexfe_hd}
\end{figure*}

\pagebreak
\bibliography{AbundanceBib.bib}
\end{document}